\documentclass[a4paper,11pt]{article}

\usepackage{jheppub}
\usepackage{amsmath}
\usepackage{amsthm}
\usepackage{amssymb}
\usepackage{graphicx}
\usepackage{epstopdf}
\usepackage{xcolor}
\usepackage{dsfont}
\usepackage{hyperref}
\usepackage{rotating}
\usepackage{afterpage}
\usepackage{arydshln}
\usepackage{cancel}
\allowdisplaybreaks

\newcommand{\be}{\begin{equation}}
	\newcommand{\ee}{\end{equation}}
\newcommand{\bea}{\begin{eqnarray}}
	\newcommand{\eea}{\end{eqnarray}}
\newcommand{\nn}{\nonumber}
\newcommand{\bdm}{\begin{displaymath}}
	\newcommand{\edm}{\end{displaymath}}

\newcommand{\dd}{\text{d}}

\DeclareMathOperator{\Ker}{Ker}

\title{Localization in Seven-Dimensional Supergravity} 

\author{Florian Gaar}
\author{and Mich\`ele Levi}

\emailAdd{florian.gaar@maths.ox.ac.uk}
\emailAdd{michele.levi@maths.ox.ac.uk}

\affiliation{Mathematical Institute, University of Oxford, \\
	Radcliffe Observatory Quarter, Woodstock Road, \\
	Oxford OX2 6GG, United Kingdom}

\abstract{
Seven-dimensional supergravity provides holographic access 
to strongly coupled six-dimensional superconformal theories 
through its highest-dimensional supersymmetric AdS vacua. 
Its unique spinor-bilinear geometry and massive higher-form dynamics 
make systematic analysis and solution-finding distinctively challenging. 
Relative equivariant localization can obtain supersymmetric observables 
in odd dimensions from fixed-point data on an auxiliary even-dimensional 
filling, without an explicit solution. 
We extend this approach to seven dimensions in maximal Cartan-gauged 
supergravity with all gauge fields active.
We formulate the Euclidean action and Killing-spinor equations in terms 
of field strengths, and construct a bilinear system adapted to the 
change in spinor reality conditions from symplectic-Majorana to Majorana. 
On the non-null gauged active spinor branch, independent, oppositely R-charged 
spinors are paired by Majorana conjugation on the same complex background. 
One algebraic gauging correction and two algebraic bilinear projections suffice 
to extend the equal-charge sector to independent Cartan charges, and the 
underlying efficient reorganization also applies in the Lorentzian theory. 
Massive 3-form self-duality fixes the quartic anomaly class 
and determines the higher-form fixed-point datum from Abelian moment maps.
Localization at the single fixed point of the filling recovers the 
general supersymmetric AdS$_7$ black-hole action, without extending the 
solution. 
We prove gauge-covariant Killing-spinor invariance under the combined 
supersymmetric Killing flow and R-symmetry action, deriving the linear 
chemical-potential constraint. 
Extremizing the constrained entropy functional and requiring real positive 
stationary entropy yields the nonlinear charge relation within the Euclidean 
analysis, equivalent to Lorentzian horizon extremality 
for the known spherical black-hole family.		
}	


\begin{document}

\maketitle
	
	
\flushbottom

\pagenumbering{arabic}
\setcounter{page}{1}

\section{Introduction}
\label{Intro}

Seven-dimensional supergravity has the highest spacetime dimension admitting 
a supersymmetric anti-de Sitter (AdS) vacuum, and thus realizes the upper 
dimensional limit of the AdS/CFT duality \cite{Maldacena:1997re}.
Indeed, the symmetries of an AdS vacuum should extend to a superconformal 
algebra on its boundary, and superconformal algebras exist only up to six 
dimensions \cite{Nahm:1977tg}.
The maximal SO$(5)$-gauged theory saturates this dimensional bound through 
its consistent $S^4$ reduction from eleven-dimensional supergravity, 
with its AdS$_7$ vacuum describing the large-$N$ dynamics 
of coincident M5-branes \cite{Maldacena:1997re,Nastase:1999cb}.
Via holography this gravitational AdS$_7$ vacuum provides access 
to the strongly interacting six-dimensional $(2,0)$ superconformal field theory 
(SCFT), for which a Lagrangian description remains unknown.
Seven-dimensional supergravity consequently supports a broad range of
holographic applications, from BPS flows \cite{Dibitetto:2017tve} and Janus 
geometries describing two-dimensional conformal defects \cite{Conti:2024rwd},  
to black holes (BHs) with spherical horizons 
or horizons constructed from Sasaki--Einstein manifolds 
\cite{Bobev:2023bxl,Bobev:2025xan}.
Among known spherical supersymmetric BH families carrying purely electric
Cartan charges in truncations of pure gauged supergravity theories,
the AdS$_7$ family shares with its AdS$_4$, AdS$_5$ analogues the richest 
macroscopic parameter space, with four independent parameters 
\cite{Hristov:2019mqp,Kunduri:2006ek,Bobev:2023bxl}.
Moreover, the seven-dimensional family supports the largest rotational sector 
with three angular momenta that accompany two electric charges.

Accessing this rich physics systematically is challenging for reasons intrinsic 
to seven dimensions, as demonstrated in \cite{Levi:2026fpa}.
The 3-form and the vector gauge fields interact through a Chern--Simons 
coupling, and with gauging the higher-form can become topologically massive, 
with its potential and field strength related by a first-order self-duality 
condition unique to seven dimensions.
Moreover, two common simplifications of the geometric spinor-bilinear system 
are simultaneously unavailable only in seven dimensions.
Middle-degree forms on the Euclidean six-dimensional base do not split 
into real self-dual and anti-self-dual parts, while the Lorentzian spinor 
requires a symplectic-Majorana doublet, yielding eight real bilinears.
Among dimensions of pure supergravity theories, the other cases 
without the simplifying middle-degree Hodge splitting have Majorana spinors 
instead and therefore yield much smaller bilinear systems.
This distinct compound complexity already holds in minimal pure ungauged 
supergravity.
The geometric description must therefore accommodate massive higher-form 
gauge interactions alongside a generic SU$(3)$ structure of a six-dimensional 
timelike base, including its defining complex 3-form, five intrinsic-torsion 
classes and numerous flux representations.
This combination of geometric intricacies and higher-form dynamics helps 
account for the difficulty of moving beyond specialized solution ans\"atze.

Furthermore, for partition functions and BH thermodynamics relevant to 
holography, these geometric structures and their description through spinor 
bilinears must also be understood in the Euclidean theory. 
Yet as we show in the following, passing to Euclidean signature reveals a 
further seven-dimensional distinction.
Among dimensions up to eleven, only in seven dimensions do the admissible 
spinor reality conditions unavoidably change from symplectic-Majorana 
to Majorana \cite{Freedman:2012zz}.
The standard Lorentzian construction using a charge-conjugate spinor pair 
therefore cannot simply be carried over.
Moreover, Euclidean supersymmetric solutions generally involve complex fields,
requiring a bilinear analysis compatible with the same complex background.

Equivariant localization offers a way to bypass much of the difficulty by 
evaluating certain supersymmetric observables from fixed-point and global 
geometric data, without requiring an explicit supergravity solution
\cite{BenettiGenolini:2023kxp,Martelli:2023oqk,BenettiGenolini:2023ndb,Colombo:2023fhu}.
The organizing symmetry is generated by a supersymmetric Killing vector 
constructed from the Killing spinors.
Supersymmetric differential relations among their bilinears link 
forms of different degrees, allowing physical integrands to be completed into 
equivariantly closed polyforms.
The Berline--Vergne--Atiyah--Bott (BVAB) fixed-point formula then enables the 
evaluation of these integrals by contributions from fixed loci of the symmetry
\cite{BerlineVergne,Atiyah:1984px}.
Remaining input consists of rotation weights, bilinear data at those loci and
appropriate global invariants.
This provides access to on-shell actions, black-hole entropies and holographic
central charges.
In particular, a Euclidean on-shell action determines the leading contribution 
of a given solution to a partition function, linking the geometric calculation 
to BH thermodynamics and protected field-theory observables.

Moreover, in odd dimensions the supersymmetric Killing vector can have 
no fixed points on spacetime, 
while Chern--Simons interactions obstruct a formulation of the
entire action as a globally gauge-invariant spacetime form.
Several extensions of localization address this setting in $d=5$
\cite{Cassani:2024kjn,Colombo:2025ihp,Cassani:2025iix,BenettiGenolini:2025icr,Colombo:2025yqy,Cassani:2026teb}. 
An earlier seven-dimensional application localized over $M_2$ in warped
AdS$_5\times M_2$ backgrounds with the 3-form gauge field set to zero
\cite{Suh:2024asy}.
Relative equivariant localization put forward in \cite{BenettiGenolini:2026cdw} 
provides a general approach to odd-dimensional settings by incorporating
the Chern--Simons couplings through an anomaly polyform on an auxiliary
even-dimensional filling of spacetime.
Together with a gauge-invariant action polyform on its bounding spacetime, 
this forms a pair whose relative closure relation cancels the boundary 
contribution in the localization formula.
The action can thus be evaluated at fixed points in the filling, which
requires extensions of the forms, but no extension of the supergravity solution.
The original construction in \cite{BenettiGenolini:2026cdw} was already 
demonstrated in five and eleven dimensions.

In this work we extend relative equivariant localization to a new odd
spacetime dimension by further developing the necessary setup in the
seven-dimensional maximal Cartan-gauged theory, with both Abelian gauge
fields and the massive 3-form active.
Starting with the Lorentzian field-strength formulation of \cite{Levi:2026fpa}, 
we analytically continue to Euclidean signature the bosonic action and the 
Killing-spinor equations (KSEs), without imposing reality conditions on the 
generic complex fields.
We work in the non-null gauged active class within the internal spinor branch 
identified in \cite{Levi:2026fpa}.
Assuming two independent Killing spinors of opposite R-charge on the
same background, we pair one with the Majorana conjugate of the other.
The resulting neutral scalar, vector and 2-form bilinears 
provide the basic bilinear system for analysis.
We first derive their equal-charge algebraic and differential relations, 
reduce the required higher-rank forms, and identify the Killing symmetry, 
gauge-field moment maps and transverse geometric identities needed for 
localization.

We assemble the relative equivariant pair by first completing 
the field strengths equivariantly and translating the massive higher-form 
self-duality into an equivariantly exact relation.
Combining this relation with the Maxwell equation and the remaining on-shell
action terms produces a global, gauge-invariant representative pair of the 
relative equivariant cohomology class.
Here the weighted closure of a transverse 4-form, tied to the geometric exact 
warp-dependent torsion 1-form of the base described in \cite{Levi:2026fpa}, 
provides the cancellation of gauging terms required for relative closure.
For spherical black holes, we subtract a reference AdS$_7$ background and
glue the two geometries into a closed 7-sphere, assuming smooth matching
of the polyforms and a compatible torus action.
We then extend the required forms to an auxiliary 8-ball filling 
and apply localization at its single fixed point.
Following the massive-form self-duality relation through this extension 
determines the quartic anomaly class and fixes the massive-form fixed-point 
datum in terms of the Abelian moment maps.
This exposes within seven dimensions the mechanism behind the quartic
structure obtained by $S^4$ integration in \cite{BenettiGenolini:2026cdw}.

To restore independent Cartan charges, we reorganize the scalar and flux terms
in the KSEs using a geometric-mean scalar and dressed symmetric and 
antisymmetric flux combinations.
A single algebraic gauging correction and two algebraic bilinear
projections then carry all the necessary additional supersymmetric input.
We use them to readily extend the bilinear relations, relative pair 
and filling construction, deriving the general two-charge, 
three-angular-momentum black-hole action of \cite{Bobev:2023bxl} 
from the seven-dimensional theory.
Our reorganization of the general KSEs also applies to the Lorentzian theory, 
providing a useful route to further applications beyond the present calculation.

Finally, we derive the restrictions that connect the localized action to
physical supersymmetric black holes, considering the supersymmetry and 
thermodynamic conditions in turn.
We first formulate the combined action of spacetime isometries and R-symmetry
on the Killing spinors and establish their gauge-covariant invariance 
along the supersymmetric Killing vector via our KSEs.
This also determines the ordinary gauge-dependent Kosmann derivative in any
chosen gauge.
We determine rotational spinor weights at fixed sets on physical spacetime
and use uniqueness of the differential KSE to propagate their relations globally.
Combining these relations with the gauge-covariant invariance yields the
linear chemical-potential constraint, without extending spinors to the
auxiliary filling.
We subsequently construct and extremize the constrained entropy functional 
at fixed conserved charges, imposing reality and positivity of its stationary 
entropy.
Comparison of the resulting additional nonlinear charge restriction with the 
known spherical-horizon BH family identifies it with Lorentzian horizon 
extremality.
These steps distinguish the supersymmetry condition from the further
thermodynamic restriction selecting the regular physical family.

The paper is organized as follows.
Section~\ref{7dcartansugra} develops the Euclidean theory and its bilinear
analysis.
Section~\ref{RelativeLocalization} constructs the relative equivariant class 
and filling, and evaluates the localized action in the equal-charge sector, 
before Section~\ref{IndependentCartanCharges} extends the analysis 
and construction to independent Cartan charges.
Section~\ref{physusysol} derives the supersymmetry and entropy conditions,
followed by our conclusions in Section~\ref{LaFin}.
Appendices~\ref{7ECliff}--\ref{EuclideanActiveHigherIdentities} provide the
Euclidean Clifford and bilinear conventions, conjugate KSEs, and the algebraic 
and higher-rank differential relations underlying the construction.

\section{7D Maximal Cartan-Gauged Supergravity}
\label{7dcartansugra}

We use the field-strength formulation of the Cartan-gauged theory 
established in \cite{Levi:2026fpa}, retaining its field definitions, 
Clifford matrices, and any other conventions and setup of \cite{Levi:2026fpa}. 
We first formulate the Euclidean theory and present the system of bosonic action 
and Killing-spinor equations (KSEs) that are needed for the relative equivariant 
localization.

\subsection{Euclidean Theory}
\label{euclidtheory}

\paragraph{Wick rotation and bosonic action.}
The bosonic fields are the metric, two scalar fields $X_I$,
two Abelian gauge fields $A_I$, with $I=1,2$, and a 3-form
gauge potential $B$.
The field strengths $F_I=\dd A_I$ and $G=\dd B$, the scalar
parametrization $X_I(\varphi_1,\varphi_2)$, and the potential
${\cal V}(X_I)$ are defined as in eqs.~(2.3)--(2.5)
of \cite{Levi:2026fpa}.
We continue the time coordinate $t=-\mathrm{i}\tau$, 
with oriented orthonormal coframes related by
$e_L^0 \mapsto -\mathrm{i} e_E^0$ and $e_L^i\mapsto e_E^i$, $i=1,\ldots,6$.
The continuation of bosonic fields is induced by the above
coordinate and coframe transformations.
The continued fields may be complex. 
From the Hodge convention in eq.~(A.3.10) of \cite{Levi:2026fpa}, which is 
retained in the Euclidean theory, we have
\be
\label{WickHodge}
\star_L \mapsto -\mathrm{i}\mskip3mu \star_E,
\qquad
\operatorname{vol}_L \mapsto -\mathrm{i} \operatorname{vol}_E.
\ee
We suppress the Euclidean subscript below. 
We extend $\star$ complex linearly and retain the kinetic terms obtained by
analytic continuation of the Lorentzian action, without complex conjugation.

Continuing the Lorentzian action $S$ in eq.~(2.2) of
\cite{Levi:2026fpa} according to the Wick-rotation prescription
$ - \mathrm{i} S\mapsto I$, we obtain the Euclidean action
\begin{align}
\label{EuclideanActionSplit}	
I&=\frac{1}{16\pi G_N}\int_M
\left({\cal L}^{\rm GI}+{\cal L}^{\rm CS}\right),
\end{align}
with the distinct parts given by
\begin{subequations}
\label{EuclideanAction}
\begin{align}
\label{EuclideanGI}	
{\cal L}^{\rm GI}
&=-\star R
+\frac12\sum_{I=1}^{2}\dd\varphi_I\wedge\star\dd\varphi_I
+\frac12\sum_{I=1}^{2}X_I^{-2}F_I\wedge\star F_I
\nn\\
&\hspace{4cm}
+\frac12(X_1X_2)^2G\wedge\star G
+2g^2\star{\cal V},
\\
\label{EuclideanCS}
{\cal L}^{\rm CS}
&= \mathrm{i} g\,B\wedge G - \mathrm{i} \mskip1mu B\wedge F_1\wedge F_2.
\end{align}
\end{subequations}
In particular, the sign of the topological mass term in eq.~\eqref{EuclideanCS} 
follows from the corresponding term in the Lorentzian action with the 
maximal supersymmetry coupling relation $m=g$ in eq.~(2.25) of 
\cite{Levi:2026fpa} already substituted in. 
With this term fixed, the anomaly 8-form is defined on an auxiliary
oriented 8-manifold $W$ with the seven-dimensional spacetime as its boundary 
$\partial W=M$, called a filling of $M$, as discussed in 
Section~\ref{RelativeLocalization}:
\be
\label{AnomalyEightForm}
{\cal P}_{(8)} \equiv \dd{\cal L}^{\rm CS}
= \mathrm{i} g\,G\wedge G - \mathrm{i} \mskip1mu G\wedge F_1\wedge F_2,
\ee
where the field strengths on the right-hand side (RHS) are also extended. 
This is the action split used in eqs.~(1)--(2) of \cite{BenettiGenolini:2026cdw}. 
Its equivariant completion will be constructed in Section~\ref{RelativeClass}.

\paragraph{Bosonic field equations and on-shell action.}
The Einstein and scalar equations of motion (EOMs) retain the 
component form of eqs.~(2.10)--(2.11) of \cite{Levi:2026fpa}, 
with the continued metric and fields. 
The Hodge factor in eq.~\eqref{WickHodge} modifies the EOMs of gauge potentials 
in eqs.~(2.12)--(2.13) of \cite{Levi:2026fpa} to
\begin{subequations}
\label{EuclideanGaugeEOM}
\begin{align}
\label{EuclideanMaxwell}	
\dd(X_I^{-2}\star F_I)&= \mathrm{i} F_J\wedge G, \qquad I,J=1,2, \quad I\neq J, 
\\*
\label{EuclideanThreeForm}
\dd\big[(X_1X_2)^2\star G\big]
&= \mathrm{i} \big(F_1\wedge F_2-2gG\big).
\end{align}
\end{subequations}
These also follow by directly varying the action in eqs.~\eqref{EuclideanAction}.
The Bianchi identities in eq.~(2.14) of \cite{Levi:2026fpa} retain their form.
The corresponding self-duality condition on the massive 3-form reads
\begin{align}
\label{EuclideanSelfDual}
(X_1X_2)^2\star G
&= \mathrm{i} \big[-2gB+\omega_{(3)}-\dd C_{(2)}\big],
&\omega_{(3)}&\equiv\frac12\big(A_1\wedge F_2+A_2\wedge F_1\big).
\end{align}
This replaces eq.~(2.15) of \cite{Levi:2026fpa}, and its exterior
derivative gives eq.~\eqref{EuclideanThreeForm}, where
the auxiliary potential $C_{(2)}$ and the combined gauge transformations in
eq.~(2.17) of \cite{Levi:2026fpa} hold here too. 

Contracting the Einstein equation in eq.~(2.10) of \cite{Levi:2026fpa} 
and substituting in eq.~\eqref{EuclideanGI} cancels 
the scalar kinetic terms and yields
\be
\label{EuclideanOnShellGI}
{\cal L}^{\rm GI}_{\rm OS}
=\frac15\sum_{I=1}^{2}X_I^{-2}F_I\wedge\star F_I
+\frac35(X_1X_2)^2G\wedge\star G
-\frac45g^2\star{\cal V}.
\ee
This part of the 7-form on-shell action is the required top component 
of the gauge-invariant polyform. 

For the equal-charge sector, we require
$A_1=A_2=A$ and $X_1=X_2=X$, as in
eqs.~(2.6)--(2.8) of \cite{Levi:2026fpa}, and
specializing eq.~\eqref{EuclideanOnShellGI} therefore gives
\begin{align}
\label{EuclideanOnShellEqual}
{\cal L}^{\rm GI}_{\rm OS}
&=\frac25X^{-2}F\wedge\star F
+\frac35X^4G\wedge\star G
-\frac45g^2\star{\cal V}.
\end{align}
The single scalar EOM in eq.~(2.11a) of \cite{Levi:2026fpa} then yields
\begin{align}
\label{EuclideanScalarEqual}
\dd\star\dd\log X
&=-\frac15X^{-2}F\wedge\star F
+\frac15X^4G\wedge\star G
+\frac{g^2}{5}X{\cal V}'(X)\star1.
\end{align}
Subtracting one half of the two flux kinetic densities from
eq.~\eqref{EuclideanOnShellEqual} and using
eq.~\eqref{EuclideanScalarEqual} gives the useful identity
\begin{align}
\label{EuclideanResidualIdentity}
{\cal L}^{\rm GI}_{\rm OS}
-\frac12\big(X^{-2}F\wedge\star F+X^4G\wedge\star G\big)
&=\frac12\dd\star\dd\log X
\nn\\
&\quad+4g^2\big(X^{-3}+2X^2\big)\star1.
\end{align}

\paragraph{Killing-spinor equations.}
Here we formulate the Euclidean field-strength KSEs as in eqs.~(2.28), (2.31) 
of \cite{Levi:2026fpa}. 
The Euclidean Clifford algebra and spinor representation are specified by 
eq.~\eqref{EuclideanClifford} in Appendix~\ref{7ECliff}.
The internal Clifford algebra Cliff$(5)$ and its real Cartan basis remain those
of Appendix~A.2 in \cite{Levi:2026fpa}.
In particular, the internal matrices $\Gamma_{\hat I}$ and 
$\Gamma_5=\Gamma_{\hat1}\Gamma_{\hat2}$ retain the real diagonal basis of 
eq.~(2.30) of \cite{Levi:2026fpa}.
Using eq.~\eqref{EuclideanGammaMap} 
the differential and the two algebraic KSEs are
\begin{align}
\label{EuclideanGravitino}
0=\bigg[&\nabla_\mu
+ \mathrm{i} g\big(A_{1\mu}\Gamma_{\hat1}+A_{2\mu}\Gamma_{\hat2}\big)
+\frac{\mathrm{i}}{4}\gamma^\nu
\big(X_1^{-1}F_{1\mu\nu}\Gamma_{\hat1}
+X_2^{-1}F_{2\mu\nu}\Gamma_{\hat2}\big)
\nn\\*
&+\frac g2(X_1X_2)^{-2}\gamma_\mu
+\frac14\gamma_\mu\gamma^\nu\partial_\nu\log(X_1X_2)
-\frac{X_1X_2}{96}\gamma_\mu{}^{\nu\lambda\sigma\rho}
G_{\nu\lambda\sigma\rho}\Gamma_5\bigg]\epsilon,
\end{align}
and
\begin{subequations}
\label{EuclideanAlgebraicKSE}
\begin{align}
\label{EuclideanAlgebraicSym}
0=\bigg[&\frac g2\big(X_1+X_2-2(X_1X_2)^{-2}\big)
-\frac58\gamma^\mu\partial_\mu\log(X_1X_2)
\nn\\*
&-\frac{\mathrm{i}}{16}\gamma^{\mu\nu}
\big(X_1^{-1}F_{1\mu\nu}\Gamma_{\hat1}
+X_2^{-1}F_{2\mu\nu}\Gamma_{\hat2}\big)
+\frac{X_1X_2}{96}\gamma^{\mu\nu\lambda\sigma}
G_{\mu\nu\lambda\sigma}\Gamma_5\bigg]\epsilon,\\*
\label{EuclideanAlgebraicAntisym}
0=\bigg[&\frac g2(X_1-X_2)
-\frac18\gamma^\mu\partial_\mu\log(X_1/X_2)
\nn\\*
&-\frac{\mathrm{i}}{16}\gamma^{\mu\nu}
\big(X_1^{-1}F_{1\mu\nu}\Gamma_{\hat1}
-X_2^{-1}F_{2\mu\nu}\Gamma_{\hat2}\big)\bigg]\epsilon.
\end{align}
\end{subequations}

In the Euclidean theory we work with complex bosonic fields 
and independent complex spinor components, similar to the formal 
prescription illustrated in \cite{Freedman:2013oja}.
In Appendix~\ref{EuclideanConjugateKSEs} we present the equations for the 
Euclidean charge conjugate $\epsilon^{c_E}$ and
Majorana conjugate $\overline\epsilon$ of the Killing spinors, defined in 
eqs.~\eqref{EuclideanChargeConjugates}, which are derived from 
eqs.~\eqref{EuclideanGravitino} and \eqref{EuclideanAlgebraicKSE}, 
as in Appendix~A.9 of \cite{Levi:2026fpa}. 
Notably, in the present Euclidean theory charge conjugation acts on the complex 
bosonic coefficients, whereas Majorana conjugation leaves these coefficients 
unchanged.
In this work, we impose no reality conditions on the Euclidean fields or spinors.
The above differential and algebraic equations provide all the spinor equations 
required for the bilinear analysis in this work.

Finally, the equal-charge KSEs follow by the same substitutions 
as in the bosonic sector, using the Cartan basis defined 
in eq.~(2.34) of \cite{Levi:2026fpa}:
\be
\label{symCartanbasis}
\Gamma_\pm \equiv (\Gamma_{\hat1}\pm\Gamma_{\hat2})/2.
\ee
The equal-charge KSEs therefore have the form of eqs.~(2.32)--(2.33) 
of \cite{Levi:2026fpa}, now with Euclidean Clifford matrices and spinors. 
In particular, the antisymmetric algebraic equation 
in eq.~(2.33b) of \cite{Levi:2026fpa} retains its simple form.

\subsection{Bilinear Analysis}
\label{BilinearAnalysis}

We first consider the equal-charge gauged active class, with $g\neq0$ and
$F\not\equiv0$ on a connected region. 
We restrict to Killing spinors whose internal components lie in $\Ker\Gamma_-$,
as discussed in Section~4.1.4 of \cite{Levi:2026fpa}.

To construct the relevant bilinears in the Euclidean theory, we first note the
following critical distinction between the Lorentzian and Euclidean Clifford 
algebras in seven dimensions.
In the Lorentzian theory, the symplectic-Majorana (SM) relation
$\chi^i=\epsilon^{ij}(\chi^j)^c$, $i=1,2$, in eq.~(2.18)
of \cite{Levi:2026fpa} relates the two opposite-R-charge spinors
$\chi^1,\chi^2$ by charge conjugation.
Together with the real bosonic background, this relation provides the
opposite-R-charge KSEs from either member of this spinor doublet.
Yet, imposing $\chi^i=\epsilon^{ij}(\chi^j)^{c_E}$ with the
Euclidean charge conjugation defined in
eqs.~\eqref{EuclideanChargeConjugates} would instead give
$\chi^i=-\chi^i$, forcing both spinors to vanish.

This is because Euclidean charge conjugation squares to $+1$, by
eq.~\eqref{EuclideanChargeSquare}, as the Euclidean spinor representation
instead admits a Majorana condition.
Among dimensions $d\leq11$, this unavoidable change from
symplectic-Majorana to Majorana reality when passing from Lorentzian 
to Euclidean signature is unique to $d=7$ \cite{Freedman:2012zz}.
Furthermore, the charge-conjugate KSEs in
Appendix~\ref{EuclideanConjugateKSEs} in the Euclidean complexified theory 
contain complex-conjugate bosonic fields, so charge conjugation does not 
generally yield the opposite-R-charge spinor solution 
on the same Euclidean background.

Moreover, a single commuting spacetime spinor, paired with its Majorana 
conjugate, would be insufficient to construct the required 1- and 2-form 
bilinears.
Eq.~\eqref{EuclideanCTranspose} makes $C\gamma_\mu$ and
$C\gamma_{\mu\nu}$ antisymmetric, so such bilinears vanish identically,
\be
\label{EuclideanSelfBilinearsVanish}
\overline\chi\gamma_\mu\chi=0,
\qquad
\overline\chi\gamma_{\mu\nu}\chi=0.
\ee

Pairing instead with its Hermitian conjugate, 
$\chi^{\dagger}=\overline{\chi^{c_E}}$, is generally nonzero. 
However, as noted above the equations for $\chi^{c_E}$ involve the 
complex-conjugate bosonic background fields, and do not in general coincide 
with the opposite-charge KSEs on the original background, as seen in
eqs.~\eqref{EuclideanGravitinoCharge}, \eqref{EuclideanAlgebraicCharge}. 
Such a pairing therefore does not in general yield bilinear
relations involving only the original complex background fields.

We instead use Majorana bilinears of two independent commuting spinors with 
opposite U$(1)$ R-charges, requiring nonzero solutions to both component KSEs 
on the same complex background. 
Since the internal spinor components lie in $\Ker\Gamma_-$,
each spinor has equal charges under the two Cartan U$(1)$ factors,
with eigenvalues $(+1,+1)$ and $(-1,-1)$ for the pair of spinors.
In the equal-charge sector, they therefore couple with opposite signs to the 
single diagonal combination $A_1+A_2=2A$.
The U$(1)$ subgroup acts diagonally on the spinor components, 
and does not itself provide one solution from the other in the Euclidean theory. 

Let us write the generic spinor in the Cartan-gauged theory in this active 
class, namely with internal components in $\Ker\Gamma_-$ \cite{Levi:2026fpa}, 
in the form
\be
\label{EuclideanActivePair}
\epsilon=\chi^1\otimes v_++\chi^2\otimes v_-,
\qquad
\Gamma_{\hat {\text{I}}} v_\pm=\pm v_\pm,\quad I=1,2.
\ee
Here $v_\pm$ are the first and fourth internal basis vectors in
eq.~(2.30) of \cite{Levi:2026fpa}, and $\chi^1,\chi^2$ solve the
corresponding spacetime components of eqs.~\eqref{EuclideanGravitino}, 
\eqref{EuclideanAlgebraicKSE}. These are precisely the components retained
in the minimal-theory doublet in eq.~(2.38) of \cite{Levi:2026fpa}.
Their opposite R-charges are under the single gauge field $A_1=A_2=A$.
Both have $\Gamma_5v_\pm=v_\pm$.

\paragraph{Scalar, 1- and 2-form bilinears.}
We therefore define, retaining the form of eqs.~(3.8a)--(3.8c) of
\cite{Levi:2026fpa}, 
\begin{subequations}
\label{EuclideanActiveBilinears}
\begin{align}
\label{EuclideanActiveScalars}
s^{ij}&\equiv\overline{\chi^i}\chi^j=s^{ji},\\
\label{EuclideanActiveVector}
K_\mu&\equiv \mathrm{i}\overline{\chi^1}\gamma_\mu\chi^2,\\
\label{EuclideanActiveTwoForm}
T_{\mu\nu}&\equiv \mathrm{i}\overline{\chi^1}\gamma_{\mu\nu}\chi^2,
\end{align}
\end{subequations}
with the gauge-invariant scalar component 
\be
\label{antsymscalar}
S_3 \equiv s^{12}.
\ee

The differential bilinear relations of the R-charged scalars generalize 
eq.~(4.50) of \cite{Levi:2026fpa} to the independent pair:
\be
\label{EuclideanChargedScalarEquations}
\dd(Xs^{11})= -4 \mathrm{i} g A\,Xs^{11},
\qquad
\dd(Xs^{22})= +4 \mathrm{i} g A\,Xs^{22}.
\ee
Similar to eqs.~(4.58)--(4.59) of \cite{Levi:2026fpa}, 
applying $\dd$ on these equations yields $gXs^{ii}F=0$. 
These equations then imply, on the connected gauged active region,
\be
\label{EuclideanChargedScalarsVanish}
s^{11}=s^{22}=0.
\ee
This is the Euclidean counterpart of Proposition~10 of
\cite{Levi:2026fpa}, corresponding to $s=0$ in the scalar
parametrization in eq.~(3.39a), with the argument in eq.~(4.59) 
now applied to each R-charged scalar independently. 

The bilinears $S_3,K,T$ defined above are gauge-invariant, as they
pair the two components of $\epsilon$ with opposite U$(1)$ R-charges.
All bilinears and their contractions are complex, without complex
conjugation in the contraction.
The Fierz identities give
\begin{subequations}
\label{EuclideanActiveFierz}	
\begin{align}
\label{Killingnorm}	
K^2 & = - S_3^2,
\\*
\iota_K T & = 0,
\end{align}
\end{subequations}
as the complex-bilinear versions of eqs.~(3.14), (3.15a) of
\cite{Levi:2026fpa}. 
We assume that the bilinear Killing vector $K$ is non-null, 
i.e.~it satisfies $K^2\neq0$ on the region considered, or equivalently 
\be
\label{nonzeroscalar}
S_3 \neq 0,
\ee
by eq.~\eqref{Killingnorm}.
This is the Euclidean counterpart of the assumption of 
a timelike Killing vector in \cite{Levi:2026fpa}.

The additional bilinear relations needed below are derived in 
Appendix~\ref{EuclideanActiveIdentities}.

\paragraph{Differential bilinear relations.}
The KSEs in eqs.~\eqref{EuclideanGravitino}, \eqref{EuclideanAlgebraicKSE}, 
together with their Majorana-conjugate versions in 
Appendix~\ref{EuclideanConjugateKSEs}, yield bilinear relations
of the same form as eqs.~(4.4b), (4.70) of \cite{Levi:2026fpa}:
\begin{subequations}
\label{EuclideanRelations}
\begin{align}
\label{EuclidTransverseDilaton}	
\iota_K \dd X & = 0,
\\
\label{EuclidKillingVector}
\nabla_{(\mu} K_{\nu)} & = 0,
\end{align}
\end{subequations}
where here $K$ and the bosonic background are generally complex.
These relations imply the invariance of $X$ and the
metric under $K$, which is used below to establish the
Killing invariance of the fields and bilinears entering the
equivariant forms.

Similarly, the Euclidean counterparts of eqs.~(4.61), (4.71), and (4.73) 
of \cite{Levi:2026fpa} read
\begin{subequations}
\label{EuclideanDifferentialBilinears}
\begin{align}
\label{EuclideanDifferentialScalar}
\dd(XS_3)&=-\iota_KF,\\*
\label{EuclideanDifferentialVector}
\dd(X^2K)&=-2XS_3F - \mathrm{i} X^4\iota_K\star G+2gX^{-2}T,\\*
\label{EuclideanDifferentialTwoForm}
\dd(X^{-2}T)&=-\iota_K G,
\end{align}
\end{subequations}
where the factor $-\mathrm{i}$ in eq.~\eqref{EuclideanDifferentialVector} follows
from the Euclidean Clifford dual in eq.~\eqref{EuclideanCliffordDual}.
The derivation of these relations is discussed 
in Appendix~\ref{EuclideanActiveIdentities}.
The Bianchi identities, together with eqs.~\eqref{EuclideanActiveFierz},
\eqref{EuclideanRelations}, and \eqref{EuclideanDifferentialBilinears}, imply
\be
\label{EuclideanKillingInvariance}
\mathcal L_KS_3=\mathcal L_KT=\mathcal L_KF=\mathcal L_KG=0.
\ee
Thus $K$ preserves the metric, the dilaton, and both field strengths.
Eqs.~\eqref{EuclideanDifferentialScalar} and
\eqref{EuclideanDifferentialTwoForm} provide the lower-degree components
of their equivariant extensions in Section~\ref{RelativeClass}.

Further higher-rank differential bilinear relations, 
which are used in our relative localization,  
are derived in Appendix~\ref{EuclideanActiveHigherIdentities} 
from the algebraic and differential bilinear relations discussed therein.

\section{Relative Equivariant Localization}
\label{RelativeLocalization}

The Chern--Simons terms in eq.~\eqref{EuclideanCS} determine the
anomaly 8-form ${\cal P}_{(8)}$ in eq.~\eqref{AnomalyEightForm}.
The relative localization principle of \cite{BenettiGenolini:2026cdw}
relates its equivariant completion on an eight-dimensional filling
to the gauge-invariant action polyform on the bounding spacetime manifold.
We first recall the setting of this statement, 
then construct the relative class directly in seven dimensions.

Let $M=\partial W$ be a closed 7-manifold bounding a compact oriented 
eight-manifold $W$, with $M$ carrying the induced boundary orientation. 
We assume a torus action on this pair $(W,M)$, preserving
the metric, scalars, and field strengths on $M$, as well as
the pair of polyforms defining the relative equivariant
cohomology class.
We recall that the Killing-vector bilinear is generally complex.
Let $V_a^W$ be the torus generators on $W$, with
$V_a=V_a^W|_M$ their restrictions to $M$.
Let $K_W$ and $K$ be induced on $W$ and $M$, such that
\be
\label{KillingVectorExtension}
K_W=\sum_a\varepsilon_a V_a^W,
\qquad
K=\sum_a\varepsilon_a V_a,
\qquad
K_W|_M=K,
\ee
with the same complex coefficients $\varepsilon_a$, 
and where $K$ is the supersymmetric Killing vector on $M$.
Define the equivariant differentials on the two manifolds by
\be
\label{RelativeCartanDifferential}
\dd_K\equiv\dd-\iota_K
\quad\text{on }M,
\qquad
\dd_{K_W}\equiv\dd-\iota_{K_W}
\quad\text{on }W,
\ee
where the two differentials are compatible with the boundary restriction,
\be
\label{RelativeCartanRestriction}
\left.(\dd_{K_W}\omega)\right|_M
=\dd_K(\omega|_M),
\ee
for any polyform $\omega$ on $W$.
The equivariant differentials satisfy the operator identities
\be
\label{RelativeCartanNilpotence}
\dd_K^2=-\mathcal L_K,
\qquad
\dd_{K_W}^2=-\mathcal L_{K_W}.
\ee
Thus $\dd_K$ and $\dd_{K_W}$ are nilpotent on the respective
Killing-invariant polyforms.

Denote the odd gauge-invariant action polyform on $M$ by $\Phi$,
and the even anomaly polyform on $W$ by $\Psi$.
They satisfy the relative closure and top-component conditions
\begin{align}
\label{RelativePairConditions}
\dd_{K_W}\Psi=0,
\qquad
\dd_K\Phi=-\Psi|_M,
\\*
\label{TopComponentConditions}
\Phi_{(7)}={\cal L}^{\rm GI}_{\rm OS},
\qquad
\Psi_{(8)}={\cal P}_{(8)},
\end{align}
so that $\Phi$ is an equivariant primitive of $-\Psi$ on $M$, 
and where we recall that the anomaly 8-form uses extended field strengths 
in eq.~\eqref{AnomalyEightForm}. 
Yet, the filling $W$ is auxiliary: it need not carry a supergravity solution.

If $K$ is nowhere zero on $M$, and the fixed-point set is written as 
\be
\mathcal Z=\{p\in W:K_W(p)=0\},
\ee 
with normal bundle $N\mathcal Z$, then the Berline--Vergne--Atiyah--Bott (BVAB) 
formula \cite{Atiyah:1984px,BerlineVergne} gives \cite{BenettiGenolini:2026cdw}
\be
\label{RelativeLocalizationFormula}
I
=\frac{1}{16\pi G_N}
\int_{\mathcal Z}\frac{\Psi|_{\mathcal Z}}{e_{K_W}(N\mathcal Z)},
\ee
where integration selects the appropriate form degree. 
For a connected fixed-point component $Z\subset\mathcal Z$,
its normal bundle formally splits into complex line bundles,
$NZ=\bigoplus_{j=1}^{r}L_j$, 
and as in eq.~(49) of \cite{BenettiGenolini:2026cdw}, 
the equivariant Euler class is then given as
\be
\label{EquivariantEulerClass}
e_{K_W}(NZ)
=\wedge_{j=1}^{r}
\left(\frac{\varepsilon_j}{2\pi}+c_1(L_j)\right),
\qquad
2r=\operatorname{codim}_W Z.
\ee
Here $\varepsilon_j\neq0$ are the weights of the linearized
$K_W$ action, normalized with respect to $2\pi$-periodic rotations,
and $c_1(L_j)$ are the first Chern classes.
The weights $\varepsilon_j$ may take complex values through
analytic continuation of the torus Lie-algebra parameters in
eq.~\eqref{RelativeLocalizationFormula}.

In this section we remain within the equal-charge gauged active class,
with the generally complex, non-null vector bilinear $K$ of
Section~\ref{BilinearAnalysis}, where we recall that the norm $K^2$
is computed without complex conjugation.
Thus due to eq.~\eqref{EuclideanActiveFierz}, $K$ has no zeros where
$S_3\neq0$, and eq.~\eqref{EuclideanKillingInvariance} ensures the 
Killing invariance required for the construction 
of the equivariant polyforms on $M$.
Applying eq.~\eqref{RelativeLocalizationFormula} 
further requires smooth global polyforms, 
and a compatible extension of the torus action and the forms to a filling $W$. 
Section~\ref{filling} constructs the spherical filling and the required 
extensions, assuming, as in \cite{BenettiGenolini:2026cdw}, smooth matching of the
polyforms across the gluing surface in the background-subtraction prescription.
This prescription provides the closed 7-manifold $M$ needed to apply 
the relative localization formula to an asymptotically AdS spacetime.

\subsection{Relative Equivariant Cohomology Class}
\label{RelativeClass}

\paragraph{Equivariant field strengths and massive 3-form self-duality.}
The scalar and 2-form differential relations in 
eqs.~\eqref{EuclideanDifferentialScalar}, \eqref{EuclideanDifferentialTwoForm}, 
fix the lower components of the equivariant field strengths:
\begin{align}
\label{Equivariant2form}
\mathcal F&\equiv F+\mu,
\qquad \mu\equiv-XS_3,
\\*
\label{Equivariant4form}
\mathcal G&\equiv G+\vartheta,
\qquad  \vartheta\equiv-X^{-2}T,
\\*
\label{EquivariantClosure}
& \dd_K\mathcal F=\dd_K\mathcal G=0,
\end{align}
where for the equivariant closure we also used the Bianchi identities 
of $F$, $G$, and eq.~\eqref{EuclideanActiveFierz}.
Here $\mu$ is the gauge-field moment map associated with $K$.
Note that here $\mathcal G$ has no 0-form component. 
Consequently, the boundary restriction of the equivariant completion of 
the anomaly 8-form in eq.~\eqref{AnomalyEightForm} is
\be
\label{BoundaryAnomalyPolyform}
\Psi|_M
=\mathrm{i}g\,\mathcal G\wedge\mathcal G
-\mathrm{i}\,\mathcal G\wedge\mathcal F\wedge\mathcal F,
\qquad \dd_K(\Psi|_M)=0.
\ee
The 8-form component is defined on the filling $W$ and vanishes upon
restriction to $M$, where $\Psi|_M$ contains only 6-, 4-, and 2-forms.

Next, we introduce the gauge-invariant polyform
\be
\label{Pdef}
P \equiv -\mathrm{i}X^4\star G+X^2K.
\ee
The 3-form EOM in eq.~\eqref{EuclideanThreeForm} combines with the
1-form bilinear relation in eq.~\eqref{EuclideanDifferentialVector}
and the Fierz identity in eq.~\eqref{EuclideanActiveFierz} 
into an equivariantly exact polyform, with $P$ its equivariant primitive:
\be
\label{EquivariantSelfDuality}
\dd_K P=\mathcal F^2-2g\mathcal G.
\ee
This equivariantly exact polyform is the equivariant form of the massive 3-form 
EOM that is underlying its self-duality. 
The equivariant primitive $P$ is already global and gauge-invariant. 
Thus the field-strength formulation
of \cite{Levi:2026fpa} carries over directly to relative localization, 
without needing to choose local representatives for the higher-form potentials 
$B$ and $C_{(2)}$ in eq.~\eqref{EuclideanSelfDual}.

\paragraph{The gauge-invariant action polyform.}
To construct the gauge-invariant action polyform $\Phi$,
we introduce two further combinations:
\begin{align}
\label{TFourForm}
\mathcal T_{(4)}& \equiv S_3^{-1}T\wedge T,
\qquad \dd_K\mathcal T_{(4)}=0,	
\\*
\label{MaxwellDescendantAndFourForm}
Q &\equiv X^{-2}\star F+\frac{\mathrm{i}}{XS_3}K\wedge T.
\end{align}
The equivariant closure in the first equation follows from 
eqs.~\eqref{EuclideanActiveFierz}, \eqref{EuclideanWeightedFourFormDerivative}.
In the Lorentzian notation of \cite{Levi:2026fpa},  
$\mathcal T_{(4)}= \mp hJ\wedge J$, so its closure is precisely the
weighted closure $\dd(hJ\wedge J)=0$ in the gauged active class, 
which is equivalent to its base having an exact torsion form
$\mathcal W_4$ depending only on the warp factor, as given in
eq.~(4.124) of \cite{Levi:2026fpa}.
This geometric property will cancel the residual gauging term in $\dd_K\Phi$.
The rank-3 differential bilinear relation in 
eq.~\eqref{EuclideanNeutralThreeFormDerivative}, together with 
the Abelian field EOM in eq.~\eqref{EuclideanMaxwell}, gives
\be
\label{EquivariantMaxwellDescendant}
\dd_K Q
=\mathrm{i}\mathcal F\wedge\mathcal G+2\mathrm{i}g\,\mathcal T_{(4)},
\ee
where the 2-form component follows from eqs.~\eqref{EuclideanActiveFierz}, 
\eqref{Equivariant2form}, \eqref{Equivariant4form}.
The additional gauging term in eq.~\eqref{EquivariantMaxwellDescendant} is
therefore fixed by the same weighted 4-form appearing in the base
geometry of \cite{Levi:2026fpa} that was noted above. 

To match the gauge-invariant on-shell action top form, 
define the odd residual polyform
$\mathcal R=\mathcal R_{(7)}+\mathcal R_{(5)}$ by
\begin{subequations}
\label{ResidualPolyform}
\begin{align}
\label{ResidualSevenForm}
\mathcal R_{(7)}
&=\frac12\dd\star\dd\log X
+4g^2(X^{-3}+2X^2)\star1,
\\
\label{ResidualFiveForm}
\mathcal R_{(5)}
&=-\frac12\iota_K(\star\dd\log X)-2gX\star T.
\end{align}
\end{subequations}
Its top component is the difference between
${\cal L}^{\rm GI}_{\rm OS}$ and one half of each flux kinetic density,
by eq.~\eqref{EuclideanResidualIdentity}.
The rank-5 differential bilinear relation in eq.~\eqref{EuclideanFiveFormIdntity} 
and eq.~\eqref{EuclidTransverseDilaton} give
\begin{align}
\label{ResidualDescentComponents}
\dd\mathcal R_{(5)}-\iota_K\mathcal R_{(7)}
&=-\mathrm{i}gF\wedge\mathcal T_{(4)},
\\*
-\iota_K\mathcal R_{(5)}
&=-\mathrm{i}g\mu\,\mathcal T_{(4)},
\end{align}
where the second equality uses 
eqs.~\eqref{EuclideanHodgeTwoForm}, \eqref{EuclideanActiveFierz}. Thus
\be
\label{ResidualEquivariantDerivative}
\dd_K\mathcal R=-\mathrm{i}g\mathcal F\wedge\mathcal T_{(4)}.
\ee
The residual polyform is therefore used to recover the gauge-invariant 
on-shell action in eq.~\eqref{EuclideanOnShellEqual},  
and cancel the gauging term left by the polyform $Q$
in eq.~\eqref{EquivariantMaxwellDescendant}.

The complete gauge-invariant action polyform is consequently
\be
\label{GaugeInvariantRelativeRepresentative}
\Phi
=\frac12\mathcal F\wedge Q
+\frac{\mathrm{i}}{2}\mathcal G\wedge P
+\mathcal R.
\ee
To verify its relative closure in eq.~\eqref{RelativePairConditions}, 
we compute the equivariant derivative of the two flux terms, obtaining
$-\Psi|_M+\mathrm{i}g\mathcal F\wedge\mathcal T_{(4)}$ by
eqs.~\eqref{EquivariantSelfDuality}, \eqref{EquivariantMaxwellDescendant}, 
and eq.~\eqref{ResidualEquivariantDerivative} cancels the remainder.
For completeness, the form components of $\Phi$ are
\begin{subequations}
\label{RelativeRepresentativeComponents}
\begin{align}
\label{RelativeRepresentativeSeven}
\Phi_{(7)}
&={\cal L}^{\rm GI}_{\rm OS},
\\*
\label{RelativeRepresentativeFive}
\Phi_{(5)}
&=-\frac12X^2T\wedge\star G
+\frac{\mathrm{i}}{2}X^2K\wedge G
-\frac{S_3}{2X}\star F
-\frac12\iota_K(\star\dd\log X)
\nn\\*
&\quad-2gX\star T
+\frac{\mathrm{i}}{2XS_3}F\wedge K\wedge T,
\\*
\label{RelativeRepresentativeThree}
\Phi_{(3)}
&=-\mathrm{i}K\wedge T,
\\*
\Phi_{(1)}&=0.
\end{align}
\end{subequations}
All components are global wherever the fields and bilinears of
Section~\ref{BilinearAnalysis} are smooth and eq.~\eqref{nonzeroscalar} holds.

We highlight that this construction of the polyforms on $M$ for the relative pair
has only used the ingredients noted in Section~\ref{7dcartansugra}, 
and the higher-rank bilinear relations 
in Appendix~\ref{EuclideanActiveHigherIdentities}, 
and that it does not require an explicit solution metric.

\subsection{Choice of Filling and Extensions}
\label{filling}

\paragraph{Extending the equivariant field strengths.}
Suppose a filling $W$ admits an equivariantly
closed extension $\mathcal F_W$ of $\mathcal F$ and a smooth
$K_W$-invariant extension $P_W$ of $P$ in eq.~\eqref{Pdef}.
The relation in eq.~\eqref{EquivariantSelfDuality} then defines
the extension of $\mathcal G$:
\begin{align}
\label{FillingFourFormExtension}
\mathcal G_W
\equiv\frac{1}{2g}\left(\mathcal F_W^2-\dd_{K_W}P_W\right),
\qquad
\mathcal G_W|_M=\mathcal G,
\qquad
\dd_{K_W}\mathcal G_W=0.
\end{align}
In particular, this construction extends forms, without
extending the Hodge star or imposing field equations on $W$.
It selects fillings for which the relation in eq.~\eqref{EquivariantSelfDuality} 
also extends. An arbitrary extension of $\mathcal G$ does not necessarily 
satisfy this relation. Extending the boundary expression in 
eq.~\eqref{BoundaryAnomalyPolyform} to $W$ gives the anomaly polyform
\be
\label{FillingAnomaly}
\Psi=\mathrm{i}g\mathcal G_W^2-\mathrm{i}\mathcal G_W\mathcal F_W^2.
\ee
Thus, eqs.~\eqref{FillingFourFormExtension}, \eqref{FillingAnomaly}, and
\eqref{GaugeInvariantRelativeRepresentative} establish all the
relative conditions in eq.~\eqref{RelativePairConditions}.

The effect of the massive 3-form self-duality is particularly transparent after
substituting eq.~\eqref{FillingFourFormExtension}
into eq.~\eqref{FillingAnomaly}:
\begin{align}
\label{QuarticAnomalyClass}
\Psi
&=-\frac{\mathrm{i}}{4g}\mathcal F_W^4+\dd_{K_W}\Lambda_W,
&
\Lambda_W&\equiv\frac{\mathrm{i}}{4g}P_W\wedge\dd_{K_W}P_W.
\end{align}
Thus the equivariant class of the anomaly polyform 
is quartic in the gauge curvature. 
If this quartic representative is used in the integral over $W$, the boundary
representative must simultaneously be shifted to $\Phi+\Lambda_W|_M$. 
This simultaneous shift preserves the relative closure conditions in
eq.~\eqref{RelativePairConditions}.
Preserving also the top-component conditions in
eq.~\eqref{TopComponentConditions} requires
$(\Lambda_W)_{(7)}|_M=0$ and $\dd(\Lambda_W)_{(7)}=0$ on $W$.
Yet, simultaneous shifts generated by smooth
$K_W$-invariant odd polyforms with no 7-form component give a
family of representatives of the same relative class satisfying
both the closure and top-component conditions.
Such shifts preserve the relative integral and the cancellation
of the BVAB boundary term displayed in eq.~(48) of
\cite{BenettiGenolini:2026cdw}.

The 0-form of $\mathcal F_W$ extends the moment map $\mu$
in eq.~\eqref{Equivariant2form} to $W$.
At an isolated fixed point $p$, set
\be 
\mu_p\equiv(\mathcal F_W)_{(0)}|_p, \qquad 
\nu_p\equiv(\mathcal G_W)_{(0)}|_p.
\ee
Regularity of $P_W$ at the fixed point and $K_W(p)=0$ then imply
\be
\label{FixedPointFourFormMoment}
\mu_p^2=2g\nu_p,
\qquad
\Psi_{(0)}|_p=-\frac{\mathrm{i}}{4g}\mu_p^4.
\ee
Note that although $\mathcal G$ has no 0-form on $M$, its extension generally
has one. 
This is the seven-dimensional counterpart of the 0-form arising
in the filling used for eleven-dimensional supergravity in
\cite{BenettiGenolini:2026cdw}.
Its fixed-point value $\nu_p$ is determined by the gauge-field moment-map value 
$\mu_p$, and hence contains no additional independent fixed-point data.

\paragraph{The spherical filling.}
For the Euclidean black holes with $S^5$ horizon in
\cite{Bobev:2023bxl}, background subtraction glues the regulated
$D^2\times S^5$ geometry to an oppositely oriented
$S^1\times D^6$ reference geometry along their common boundary $S^1\times S^5$.
With the standard gluing, the closed manifold and its filling are
\be
\label{SphericalFilling}
M\cong S^7,
\qquad W=D^8.
\ee
This is the seven-dimensional factor of the filling used in
\cite{BenettiGenolini:2026cdw}. The use of $\mathbb R^8$ in its
fixed-point calculation therein describes the interior of the ball, 
which shares the same fixed-point data: the compact filling in 
eq.~\eqref{SphericalFilling} is the one with boundary $S^7$.

Here and below the on-shell action is understood in that background-subtraction
prescription, with the polyforms $\Phi$, $\mathcal F$, and $P$ 
matched smoothly across the regulated gluing.
The glued space is an auxiliary integration cycle, and is not
required to be a regular supergravity solution. 
The relative closure in eq.~\eqref{RelativePairConditions} 
does not by itself establish cancellation of all asymptotic boundary
terms in an arbitrary holographic renormalization scheme.
For the BH solutions with an $S^5$ horizon, the comparison below uses the scheme 
of Section~3.4 of \cite{Bobev:2025xan}, 
in which the AdS$_7$ vacuum has a vanishing on-shell action.

On $W=D^8\subset\mathbb C^4$, the torus acts by independent rotations
of the four complex coordinate planes, so that
$V_a^W=\partial_{\phi_a}$ in eq.~\eqref{KillingVectorExtension}:
\be
\label{BallTorusWeights}
K_W=\sum_{a=0}^{3}\varepsilon_a\partial_{\phi_a},
\qquad \phi_a\sim\phi_a+2\pi.
\ee
For nonzero weights it has a single fixed point at the origin 
and no fixed points on the boundary $M=S^7$. 
The ball is smooth and admits a spin structure. 
We choose the plane orientations defining the sign conventions for
the weights in eq.~\eqref{EquivariantEulerClass} so that their 
product orientation induces the prescribed orientation on $M=\partial W$.

Let us now construct the required form extensions assumed above,
explicitly on this filling.
Since the second de Rham cohomology group is trivial, $H^2(S^7)=0$,  
every closed 2-form on $S^7$ is exact, so the field strength
$F=\dd A$ admits a global 1-form potential.
Then its global potential is also Killing-invariant 
up to a gauge transformation,
\be
\label{compenstaing}
\lambda\equiv\mu+\iota_KA,\qquad
\dd\mu=\iota_KF 
\qquad \Longrightarrow \qquad
\mathcal L_K A=\dd\lambda,
\ee
where $\mu$ is the moment map from eq.~\eqref{Equivariant2form},
and we have used eq.~\eqref{EuclideanDifferentialScalar}.
Since both $\mu$ and $A$ are global, $\lambda$ is a globally defined scalar.
Moreover, since $F$ is torus-invariant, averaging the global potential
$A$ over the periodic parameters of the torus action gives a global 
torus-invariant primitive $\hat A$ of $F$. 
Since $\hat A$ is Killing-invariant, eq.~\eqref{compenstaing} 
then gives for the corresponding compensating scalar
\be
\label{BoundaryLiftConstant}
\hat\lambda = \mu+\iota_K\hat A = \text{constant},
\ee
which 
is also globally defined.
The equivariant field strength can then be written as
\be
\label{BoundaryEquivariantLift}
\mathcal F=\hat\lambda+\dd_K\hat A.
\ee

We use the auxiliary potential $\hat A$ to construct the extension,
without fixing the gauge of the original $A$.
To extend $\hat A$ and $P$ into the punctured ball, we pull them
back along the radial projection onto $S^7$ and multiply the
pullbacks by a smooth radial function equal to 1 near the boundary
and 0 near the centre.
This gives smooth $K_W$-invariant extensions $\hat A_W$ and $P_W$
with the prescribed boundary values. We then set
\be
\label{ExplicitBallExtension}
\mathcal F_W=\hat\lambda+\dd_{K_W}\hat A_W.
\ee
The constancy of $\hat\lambda$ and the $K_W$-invariance of $\hat A_W$
ensure equivariant closure of $\mathcal F_W$ by
eq.~\eqref{RelativeCartanNilpotence}.
Since $\hat A_W|_M=\hat A$, eqs.~\eqref{RelativeCartanRestriction}
and \eqref{BoundaryEquivariantLift} give
$\mathcal F_W|_M=\mathcal F$.
Since $\hat A_W$ vanishes near the origin, the 0-form component
of eq.~\eqref{ExplicitBallExtension} gives
\be
\label{BallFixedPointMoment}
\mu_0\equiv(\mathcal F_W)_{(0)}|_{p=0}=\hat\lambda.
\ee
The extensions $\mathcal F_W$ and $P_W$ constructed above determine
the equivariantly closed extension $\mathcal G_W$ of $\mathcal G$
through eq.~\eqref{FillingFourFormExtension}.

As noted in eq.~\eqref{BoundaryLiftConstant}, 
Killing invariance of $\hat A$ makes $\hat\lambda$ constant.
The particular Killing gauge in eq.~(4.64) of \cite{Levi:2026fpa}
is fixed by the differential relation following from
the KSEs and would additionally set $\lambda=0$ for our spinors.
We do not impose this stronger condition on $\hat A$.
For any regular filling potential, the contraction with $K_W$ vanishes
at the fixed point.
For $\mu_0\neq0$, eq.~\eqref{BallFixedPointMoment} therefore shows that
a gauge with $\lambda=0$ would be singular at the fixed point.
Therefore, as the fixed-point value $\Psi_{(0)}|_{p=0}$ is proportional to
$\mu_0^4=(\hat\lambda)^4$ by eq.~\eqref{FixedPointFourFormMoment}, we see that
a nonzero action in eq.~\eqref{RelativeLocalizationFormula} is associated with 
a regular filling gauge, in which $\hat\lambda\neq0$.

\subsection{Localization and Black-Hole Action}
\label{localizedaction}

To evaluate the relative localization formula in
eq.~\eqref{RelativeLocalizationFormula}, we use the equivariant Euler
class in eq.~\eqref{EquivariantEulerClass} for the tangent space at the
single isolated fixed point at the filling's origin,
\begin{align}
\label{SingleFixedPointAction}
e_{K_W}(T_0W)&=\frac{\varepsilon_0\varepsilon_1
\varepsilon_2\varepsilon_3}{(2\pi)^4},
\end{align}
which yields the following Euclidean on-shell action:
\begin{align}
\label{OSaction}
I&=-\frac{\mathrm{i}\pi^3}{4G_Ng}
\frac{\mu_0^4}{\varepsilon_0\varepsilon_1
\varepsilon_2\varepsilon_3}.
\end{align}
The result is invariant under a common rescaling of $K_W$, since
$\mu_0$ and all four tangent weights scale together.
The $S^4$ reduction of the classical theory in
\cite{BenettiGenolini:2026cdw} produces the same quartic dependence
from the M5-brane anomaly polynomial. Here it follows directly from
the seven-dimensional equation of the massive 3-form.

To compare this result with the black-hole action in \cite{Bobev:2023bxl}, 
let $b$ be the common R-symmetry weight in the two SO$(5)$ Cartan planes.
The internal spinor has half the vector weight in each plane, while
eq.~\eqref{EuclideanGravitino} couples it through
$\mathrm{i}gA_I\Gamma_{\hat I}$.
Thus, with the internal plane orientations chosen to match the Cartan
generators $\Gamma_{\hat I}$, and using $\mu_0=\hat\lambda$ from
eq.~\eqref{BallFixedPointMoment}, we identify
\be
\label{CartanWeightNormalization}
b=2g\mu_0.
\ee
We also record the equal-charge specialization of the spinor-weight relation
in eq.~(35) of
\cite{BenettiGenolini:2026cdw}:
\be
\label{FillingSpinorWeightConstraint}
2b+\sum_{a=0}^{3}\varepsilon_a=0.
\ee
In Section~\ref{susycond}, we derive this constraint from the KSEs and
the torus action on the physical spacetime, by establishing Killing-spinor
invariance under the combined spacetime rotations and R-symmetry
transformations. 
Let us highlight that this derivation uses the topology of physical
spacetime and is independent of the auxiliary gluing and filling used
in the localization.

For this comparison, we first denote the positive coupling in
\cite{Bobev:2023bxl,Bobev:2025xan} by $\mathfrak g$.
Their Lorentzian topological mass term corresponds to ours under
\be
\label{BlackHoleCouplingMap}
g=-\mathfrak g,
\qquad \mathfrak g>0.
\ee
This is the bosonic convention map discussed in Appendix~B of 
\cite{Bobev:2023bxl} and in Section~2 and Appendix~A.16 of \cite{Levi:2026fpa}.
We then define the dimensionless chemical potentials $\Delta$ and $\omega_i$,
using the inverse temperature $\beta$,
common electrostatic potential $\Phi_{\rm e}$,
and angular velocities $\Omega_i$ of \cite{Bobev:2023bxl}:
\be
\label{BlackHoleChemicalPotentials}
\Delta\equiv-2\mathfrak g\beta(\Phi_{\rm e}-1),
\qquad
\omega_i\equiv\beta(\Omega_i+\mathfrak g).
\ee
Using the sign conventions for the spacetime and internal rotation weights
and second-sheet branch of
\cite{BenettiGenolini:2026cdw}, we identify
\be
\label{BlackHoleWeightDictionary}
b=-\Delta, 
\qquad \varepsilon_0=2\pi\mathrm{i},
\qquad \varepsilon_i=\omega_i, 
\quad i=1,2,3.
\ee
Under these identifications, the weight relation in
eq.~\eqref{FillingSpinorWeightConstraint} becomes
\be
\label{EqualChargeChemicalConstraint}
2\Delta-\omega_1-\omega_2-\omega_3=2\pi\mathrm{i}.
\ee
This is the equal-charge version of eq.~(5.2) of \cite{Bobev:2023bxl}.
Choosing the conjugate sheet changes the sign on the RHS together with the
corresponding weight data.

Substituting eqs.~\eqref{CartanWeightNormalization},
\eqref{BlackHoleCouplingMap}, \eqref{BlackHoleWeightDictionary} into
eq.~\eqref{OSaction} yields
\be
\label{LocalizedEqualChargeAction}
I_{\rm BH}
=\frac{\pi^2}{128G_N\mathfrak g^5}
\frac{\Delta^4}{\omega_1\omega_2\omega_3}
=\frac{N^3}{24}\frac{\Delta^4}{\omega_1\omega_2\omega_3},
\qquad
N^3=\frac{3\pi^2}{16G_N\mathfrak g^5}.
\ee
This agrees with the equal-charge specialization of eqs.~(4.33),
(5.3)--(5.6) of \cite{Bobev:2023bxl}, and with the independently
renormalized action in \cite{Bobev:2025xan} in its scheme with zero
vacuum action.
We highlight that the fourth power in the numerator is fixed by the massive
self-duality of the 3-form used in the equivariant class of the anomaly polyform
in eq.~\eqref{QuarticAnomalyClass}.
This comparison concerns the complex supersymmetric Euclidean solutions.
The additional entropy condition is derived within Euclidean black-hole
thermodynamics in Section~\ref{RealMaxEntropy}, where we then compare it
with the Lorentzian horizon extremality of \cite{Bobev:2023bxl}.

\section{Independent Cartan Charges}
\label{IndependentCartanCharges}

The equal-charge analysis and localization can be extended 
to independent Cartan charges 
using the symmetric and antisymmetric KSEs in eqs.~(2.28), (2.31) 
of \cite{Levi:2026fpa}, whose Euclidean counterparts are in 
eqs.~\eqref{EuclideanGravitino}, \eqref{EuclideanAlgebraicKSE}.
We thus retain the corresponding $\Gamma_\pm$ Cartan basis 
in eq.~\eqref{symCartanbasis} from Appendix~A.2 of \cite{Levi:2026fpa}.
Building on this formulation, a change of variables 
replacing the equal-charge dilaton by the geometric mean of the two 
dilatons, and the field strength by a symmetric weighted sum of the 
two field strengths, allows us to recover 
a significant equal-charge baseline.
Only the algebraic KSEs contribute unequal-charge deviations, with
a single gauging correction term in the symmetric equation, 
and two additional bilinear projections from the antisymmetric one.
We use this critical observation to extend the bilinear analysis
and relative equivariant pair of Sections~\ref{BilinearAnalysis} and
\ref{RelativeClass}, respectively, while retaining the filling 
constructed in Section~\ref{filling}.

\subsection{Bilinear Relations}
\label{BilinearRelations}

We retain the pair of independent, oppositely R-charged Euclidean spinors 
and bilinears introduced in Section~\ref{BilinearAnalysis} and defined in
eqs.~\eqref{EuclideanActivePair}, \eqref{EuclideanActiveBilinears}, 
respectively, and define new variables
\be
\label{CartanDressedSources}
\mathcal X \equiv\sqrt{X_1X_2},
\quad x \equiv\sqrt{\frac{X_1}{X_2}},
\quad \qquad F_\pm \equiv\frac12(x^{-1}F_1\pm xF_2),
\ee
with the geometric mean $\mathcal X$ and the scalar ratio in $x$, 
and we have $X_1=\mathcal Xx$, $X_2=\mathcal X/x$.

We work on the same non-null gauged active spinor branch $\Ker \Gamma_-$ 
with $K^2\neq0$ and $S_3\neq0$. 
We recall that in the $\Gamma_\pm$ basis, the pair of spinors obeys 
$\Gamma_-\epsilon=0$. 
In the equal-charge limit,
\begin{align}
\label{CartanEqualChargeLimit}
\mathcal X&=X,\qquad x=1, \\*
F_+&=F, \qquad F_-=0,
\end{align}
and the antisymmetric KSE vanishes on this pair of spinors, 
consistently with eq.~(2.33b) of \cite{Levi:2026fpa}.

With unequal charges, the scalar and field-strength terms of the differential 
KSE in eq.~\eqref{EuclideanGravitino} still take their equal-charge form under
\be
\label{CartanSourceSubstitution}
X\longmapsto\mathcal X,\qquad F\longmapsto F_+.
\ee
The charged connection becomes $g(A_1+A_2)$, and thus it drops out 
from the differential relations of our R-neutral bilinears in 
eqs.~\eqref{EuclideanActiveBilinears}. 
The argument for the vanishing R-charged scalar bilinears also extends. 
Their differential equations read
\begin{align}
\label{CartanChargedScalarEquations}
\dd(\mathcal Xs^{11})
=-2\mathrm{i}g(A_1+A_2)\mathcal Xs^{11}, \qquad
\dd(\mathcal Xs^{22})
=+2\mathrm{i}g(A_1+A_2)\mathcal Xs^{22},
\end{align}
and as in \cite{Levi:2026fpa}, we similarly get
$g\mathcal Xs^{ii}(F_1+F_2)=0$,
thus a net nonvanishing flux $F_1+F_2\not\equiv0$ on a connected region implies 
eq.~\eqref{EuclideanChargedScalarsVanish} here too.
With $s^{11}=s^{22}=0$ and $S_3\neq0$, the Fierz and Hodge identities
in Appendix~\ref{EuclideanActiveIdentities} remain unchanged.

Using the map in eq.~\eqref{CartanSourceSubstitution},
the symmetric algebraic KSE extends as
\be
\label{CartanSymmetricCorrection}
\eqref{EuclideanAlgebraicSym}[X,F] = 0
\quad \longmapsto \quad 
\eqref{EuclideanAlgebraicSym}[\mathcal X,F_+]
+g\frac{\mathcal X}{2}(x+x^{-1}-2)\epsilon = 0,
\ee
hence its new second gauging term is the only correction to the 
map in eq.~\eqref{CartanSourceSubstitution}.

More notably, under the map in eq.~\eqref{CartanSourceSubstitution},
the antisymmetric KSE extends as
\be
\label{CartanFlavourOperator}
\begin{aligned}
\underbrace{\eqref{EuclideanAlgebraicAntisym}[X,F]}_{0} = 0
\quad \longmapsto\quad{}&
\underbrace{\eqref{EuclideanAlgebraicAntisym}[\mathcal X,F_+]}_{0}
\\
& + \left[g\frac{\mathcal X}{2}(x-x^{-1})
-\frac14\gamma^\mu\partial_\mu\log x
-\frac{\mathrm{i}}{8\mathcal X}\gamma^{\mu\nu}F_{-\mu\nu}\Gamma_+
\right]\epsilon = 0.
\end{aligned}
\ee
Its bilinear projections yield two new independent relations, written
as a 1-form and a 4-form,
\begin{subequations}
\label{CartanFlavourProjections}
\begin{align}
\label{nonequal1formrelation}	
\iota_KF_-&=-\frac{\mathcal X}{x} S_3\dd x\,,
\\*
\iota_K\star F_-
&=-\mathrm{i}(F_-)_{\perp}\wedge T-\mathrm{i}g\frac{\mathcal X^2}{S_3}(x-x^{-1})T\wedge T,
& \quad (F_-)_{\perp}&\equiv F_-+S_3^{-2}K\wedge\iota_KF_-,
\end{align}
\end{subequations}
where the first relation follows by combining the rank-1 projections.
Using this relation, we rewrite the transverse rank-2 projection as the
second relation by applying $\iota_K\star$, using 
the geometric identities of higher-rank bilinears 
in eqs.~\eqref{EuclideanMixedReductions} and the Hodge identities in
Appendix~\ref{EuclideanActiveIdentities}.
This amounts to Hodge duality on the six-dimensional space transverse
to $K$, with the conversion invertible since $K$ is non-null.
The 4-form presentation provides the wedge-product identity needed for the 
higher-rank differential relations of the individual Cartan factors in 
eqs.~\eqref{CartanHigherIdentities}, \eqref{CartanFiveFormIdentity} below.
Note that the first relation implies $\iota_K\dd x=0$, 
and the second one implies $T\lrcorner F_-=-2g\mathcal X^2S_3(x-x^{-1})$, 
which account for the scalar projections.

\textit{The algebraic gauging correction in 
eq.~\eqref{CartanSymmetricCorrection}
and the two new algebraic bilinear relations in 
eqs.~\eqref{CartanFlavourProjections} 
constitute all the additional supersymmetric input required 
to activate the two Cartan charges independently.} 

With the map in eq.~\eqref{CartanSourceSubstitution},
the differential bilinear relations extend readily,
and using eq.~\eqref{nonequal1formrelation}, 
the scalar one additionally gives the individual Cartan relations, with the 
second substitution below applied separately to each Cartan factor, $I=1,2$:
\begin{subequations}
\label{CartanDifferentialBilinears}
\begin{align}
\label{EuclideanDifferentialBilinearsUnequal}	
\eqref{EuclideanDifferentialBilinears}[X,F]
& \quad \longmapsto \quad
\eqref{EuclideanDifferentialBilinears}[\mathcal X,F_+],
\\*
\label{diffindiv}
\eqref{EuclideanDifferentialScalar}[\mathcal X,F_+]
&\quad \Longrightarrow \quad 
\eqref{EuclideanDifferentialScalar}[X_I,F_I],
\qquad I=1,2.
\end{align}
\end{subequations}
The Killing-vector property and Killing invariance follow by the same arguments 
of Section~\ref{BilinearAnalysis}. 
In particular, eq.~\eqref{EuclideanWeightedFourFormDerivative} 
still gives the rank-4 relation in eq.~\eqref{TFourForm}.

Finally, using the map in eq.~\eqref{CartanSourceSubstitution} and the correction
in eq.~\eqref{CartanSymmetricCorrection}, the rank-3 and -5 relations in 
Appendix~\ref{EuclideanActiveHigherIdentities} extend as
\begin{align}
\label{CartanInheritedHigherIdentities}
\eqref{EuclideanNeutralThreeFormDerivative}[X,F]
& \quad \longmapsto \quad
\eqref{EuclideanNeutralThreeFormDerivative}[\mathcal X,F_+]
+g(x+x^{-1}-2)\mathcal T_{(4)},
\\*
\label{CartanInheritedFiveFormIdentity}
\eqref{EuclideanFiveFormIdntity}[X,F]
& \quad \longmapsto \quad 
\eqref{EuclideanFiveFormIdntity}[\mathcal X,F_+]
-2g\mathcal X^2(x+x^{-1}-2)\star K.
\end{align}
The new bilinear relations in eqs.~\eqref{CartanFlavourProjections} and
the unchanged Hodge identities in eqs.~\eqref{EuclideanActiveHodge} 
imply the additional algebraic constraints
\begin{align}
\label{CartanFlavourCorrections}
\frac{1}{S_3}\dd\log x\wedge K\wedge T
&=\mathrm{i}\mathcal X^{-1}\iota_K\star F_-
-\mathcal X^{-1}F_-\wedge T-g\mathcal X(x-x^{-1})\mathcal T_{(4)},
\\*
\label{CartanFlavourFiveFormCorrection}
\dd\log x\wedge\star T
&=\frac{\mathrm{i}}{2\mathcal X}F_-\wedge\mathcal T_{(4)}
+2g\mathcal X(x-x^{-1})\star K.
\end{align}
Combining
eqs.~\eqref{CartanInheritedHigherIdentities}--\eqref{CartanFlavourFiveFormCorrection}
gives the individual Cartan differential relations
\begin{align}
\label{CartanHigherIdentities}
\dd\left(\frac{K\wedge T}{X_IS_3}\right)
&=-\mathrm{i}X_I^{-2}\iota_K\star F_I
-\mathcal X^{-2}F_J\wedge T-X_JS_3G+2g\mathcal T_{(4)}, 
\quad I,J=1,2, \quad I\neq J,
\\*
\label{CartanFiveFormIdentity}
\dd(X_I\star T)
&=\frac{\mathrm{i}}{2}F_I\wedge\mathcal T_{(4)}
-2g\left(X_I\mathcal X^{-4}+2\mathcal X^2\right)\star K,
\qquad I=1,2.
\end{align}

\subsection{Relative Equivariant Pair}
\label{RelativePair}

The construction in Section~\ref{RelativeClass} also extends readily. 
We use the common variables $(\mathcal X,F_+)$ and the individual
Cartan pairs $(X_I,F_I)$ as complementary descriptions, 
with sums and products formed only after applying the substitution separately 
to each Cartan factor.
The definitions in eqs.~\eqref{Equivariant2form},
\eqref{MaxwellDescendantAndFourForm}, and
eqs.~\eqref{Equivariant4form}, \eqref{Pdef}
extend, respectively, as follows, with the first substitution
applied separately to each Cartan factor, $I=1,2$:
\begin{align}
\label{CartanEquivariantCurvatures} 
(\mathcal F,\mu,Q)[X,F]
& \quad \longmapsto \quad 
(\mathcal F_I,\mu_I,Q_I)[X_I,F_I],
\qquad I=1,2,
\\*
\label{CartanDescendants}
(\mathcal G,P)[X]
& \quad \longmapsto \quad 
(\mathcal G,P)[\mathcal X].
\end{align}
Using also eqs.~\eqref{CartanDifferentialBilinears}, the equivariant closure in 
eq.~\eqref{EquivariantClosure} also carries over under these maps.
Using eqs.~\eqref{EuclideanGaugeEOM},
\eqref{CartanDifferentialBilinears}, \eqref{CartanHigherIdentities},
and eq.~\eqref{EuclideanActiveFierz}, 
the equivariant relations extend as follows, with the second
substitution applied separately for the two ordered pairs
$(I,J)=(1,2)$ and $(2,1)$:
\begin{align}
\label{CartanEquivariantDescents}
\eqref{EquivariantSelfDuality}:&\qquad
\mathcal F^2 \quad \longmapsto \quad \mathcal F_1\wedge\mathcal F_2,
\\*
\label{CartanEquivariantMaxwellDescendant}
\eqref{EquivariantMaxwellDescendant}:&\qquad
(Q,\mathcal F) \quad \longmapsto \quad (Q_I,\mathcal F_J),
\quad I,J=1,2,\quad I\neq J.
\end{align}

The residual polyform $\mathcal R=\mathcal R_{(7)}+\mathcal R_{(5)}$ in 
eqs.~\eqref{ResidualPolyform} extends as
\be
\label{CartanResidualPolyform}
\mathcal R[X] \quad \longmapsto \quad
\mathcal R[\mathcal X]
+g(x+x^{-1}-2)
\left(2g\mathcal X^{-3}\star1-\mathcal X\star T\right).
\ee
The sum of two 5-form identities in eq.~\eqref{CartanFiveFormIdentity} extends 
eq.~\eqref{ResidualEquivariantDerivative}, which is linear in $\mathcal F$, 
under $\mathcal F\mapsto\frac12(\mathcal F_1+\mathcal F_2)$.
The scalar EOM in eq.~\eqref{EuclideanScalarEqual} extends as
\be
\label{CartanProductScalarEOM}
\eqref{EuclideanScalarEqual}[X,F]
\quad  \longmapsto \quad 
\eqref{EuclideanScalarEqual}[\mathcal X,F_+]
-\frac15\mathcal X^{-2}F_-\wedge\star F_- 
+ \frac{12}{5} g^2 \mathcal X^{-3} (x+x^{-1}-2)\star1.
\ee
Together with eqs.~\eqref{EuclideanOnShellGI},
\eqref{CartanResidualPolyform}, this extends
eq.~\eqref{EuclideanResidualIdentity} to
\be
\label{CartanResidualActionIdentity}
\mathcal R_{(7)}
={\cal L}^{\rm GI}_{\rm OS}
-\frac14\sum_I X_I^{-2} F_I\wedge \star F_I
-\frac12\mathcal X^4 G\wedge \star G.
\ee

With the extensions of $\mathcal G$, $P$, and $\mathcal R$ above,
the relative pair in eqs.~\eqref{GaugeInvariantRelativeRepresentative},
\eqref{BoundaryAnomalyPolyform} extends as
\begin{align}
\label{CartanRelativeRepresentative}
\Phi[\mathcal F\wedge Q]
& \quad \longmapsto \quad
\Phi\left[\frac12\sum_I\mathcal F_I\wedge Q_I\right],
\\*
\label{CartanRelativeConditions}
\left.\Psi[\mathcal F^2]\right|_M
& \quad \longmapsto \quad 
\left.\Psi[\mathcal F_1\wedge\mathcal F_2]\right|_M.
\end{align}
The resulting pair satisfies the relative closure in 
eq.~\eqref{RelativePairConditions} by
eqs.~\eqref{CartanEquivariantDescents}--\eqref{CartanEquivariantMaxwellDescendant} 
and the extension of eq.~\eqref{ResidualEquivariantDerivative}, and the 
gauge-invariant top-component condition in eq.~\eqref{TopComponentConditions} by 
eq.~\eqref{CartanResidualActionIdentity}.

\subsection{Extensions and Localization}
\label{FullCartanExtensions}

We retain the background-subtraction prescription, the auxiliary filling
$D^8$ with boundary $S^7$, and the orientations and tangent weights of
Sections~\ref{filling}, \ref{localizedaction}.
The construction in
eqs.~\eqref{BoundaryLiftConstant}--\eqref{BallFixedPointMoment}
applies separately to $\mathcal F_1$ and $\mathcal F_2$.
Their Killing-invariant global auxiliary potentials $\hat A_I$
define constants $\hat\lambda_I$, which give the respective moment-map
values at the ball's fixed point:
\be
\label{independentmoments}
\hat\lambda_I=\mu_I+\iota_K\hat A_I, \qquad \mu_{I,0}=\hat\lambda_I.
\ee
We extend $P$ using the same invariant radial cutoff and set
\be
\label{CartanFourFormExtension}
\Psi=\mathrm{i}g\mathcal G_W^2
-\mathrm{i}\mathcal G_W\wedge\mathcal F_{1,W}\wedge\mathcal F_{2,W}, 
\qquad \dd_{K_W}P_W = \mathcal F_{1,W}\wedge\mathcal F_{2,W} - 2 g \mathcal G_W.
\ee
Eq.~\eqref{CartanEquivariantDescents} ensures boundary matching.
The extension and localization of Sections~\ref{filling}, \ref{localizedaction}
therefore require only the replacement
$\mathcal F_W^2\mapsto\mathcal F_{1,W}\wedge\mathcal F_{2,W}$.
In particular, eq.~\eqref{QuarticAnomalyClass} becomes
\be
\label{CartanQuarticAnomaly}
\Psi=-\frac{\mathrm{i}}{4g}
(\mathcal F_{1,W}\wedge\mathcal F_{2,W})^2
+\frac{\mathrm{i}}{4g}\dd_{K_W}\left(P_W\wedge\dd_{K_W}P_W\right).
\ee
The corresponding simultaneous boundary shift of $\Phi$ is precisely the one
explained below eq.~\eqref{QuarticAnomalyClass}. 

Equivalently, regularity of $P_W$ at the fixed point gives 
$2g(\mathcal G_W)_{(0)}|_0=\mu_{1,0}\mu_{2,0}$, hence
$\Psi_{(0)}|_0=-\mathrm{i}\mu_{1,0}^2\mu_{2,0}^2/(4g)$, similar to 
eq.~\eqref{FixedPointFourFormMoment}.
Using eq.~\eqref{SingleFixedPointAction} fixes the action as 
\be
\label{CartanDirectLocalizedAction}
I=-\frac{\mathrm{i}\pi^3}{4G_Ng}
\frac{\mu_{1,0}^2\mu_{2,0}^2}
{\varepsilon_0\varepsilon_1\varepsilon_2\varepsilon_3}.
\ee

To compare this localized result with the black-hole action, we identify
the weight in each Cartan plane from its spinor charge in
eq.~\eqref{EuclideanGravitino}, as in the argument for
eq.~\eqref{CartanWeightNormalization}:
\be
\label{IndependentCartanWeights}
b_I=2g\mu_{I,0}.
\ee
We also record the general spinor-weight relation, whose equal-charge
specialization is eq.~\eqref{FillingSpinorWeightConstraint}:
\be
\label{IndependentCartanWeightsRelation}
\sum_{I=1}^2 b_I+\sum_{a=0}^3\varepsilon_a=0.
\ee
In Section~\ref{susycond}, we derive this constraint from the KSEs
and the torus action on the physical spacetime, independently of the
auxiliary gluing and filling used in the localization.

With $g=-\mathfrak g$, we extend the dictionary in
eq.~\eqref{BlackHoleWeightDictionary} by
\be
\label{IndependentWeightDictionary}
\Delta_I=-2\mathfrak g\beta(\Phi_{{\rm e},I}-1),
\quad b_I=-\Delta_I,
\qquad\varepsilon_0=2\pi \mathrm{i},
\quad\varepsilon_i=\omega_i=\beta(\Omega_i+\mathfrak g).
\ee
The electrostatic normalization is that of eqs.~(5.4)--(5.5) of
\cite{Bobev:2023bxl}. We thus obtain the weight relation
\be
\label{IndependentChemicalConstraint}
\Delta_1+\Delta_2-\omega_1-\omega_2-\omega_3=2\pi \mathrm{i},
\ee	
and the unequal-charge black-hole action
\be
\label{IndependentCartanAction}
I_{\rm BH}
=\frac{\pi^2}{128G_N\mathfrak g^5}
\frac{\Delta_1^2\Delta_2^2}{\omega_1\omega_2\omega_3}
=\frac{N^3}{24}
\frac{\Delta_1^2\Delta_2^2}{\omega_1\omega_2\omega_3}.
\ee
Here the relation between $N$, $G_N$, and $\mathfrak g$ is
unchanged from eq.~\eqref{LocalizedEqualChargeAction}.
These results agree with eqs.~(4.33), (5.2)--(5.6) of
\cite{Bobev:2023bxl}. Eqs.~(3.27)--(3.28) of
\cite{Bobev:2025xan} give the corresponding holographically
renormalized comparison in the scheme in which the AdS$_7$ vacuum 
has a vanishing on-shell action. 
As in Section~\ref{localizedaction}, this comparison concerns complex
supersymmetric Euclidean solutions.
The additional entropy condition is derived within Euclidean black-hole
thermodynamics in Section~\ref{RealMaxEntropy}.

\section{Physical Supersymmetric Solutions}
\label{physusysol}

The localized action in eq.~\eqref{CartanDirectLocalizedAction} provides
the starting point for studying general supersymmetric Euclidean solutions
with the topology and toric symmetry specified in Section~\ref{filling}.
In the present section, we proceed to gradually constrain 
these general solutions to physical supersymmetric Euclidean ones. 
In Section~\ref{susycond} we first establish, in a gauge-covariant formulation, 
the invariance of our Killing spinors under 
the combined supersymmetric Killing isometry and R-symmetry action, and derive 
the supersymmetry constraint on their spacetime and R-symmetry weights.
We thus establish the linear relation among these weights, completing 
the independent derivation of eqs.~\eqref{CartanWeightNormalization}, 
\eqref{FillingSpinorWeightConstraint}, \eqref{IndependentCartanWeights},  
\eqref{IndependentCartanWeightsRelation} deferred in 
Sections~\ref{localizedaction}, \ref{FullCartanExtensions}.
In the known six-parameter black-hole family, this supersymmetry
constraint selects a five-parameter subset \cite{Bobev:2023bxl}.
Then in Section~\ref{RealMaxEntropy} we construct the entropy functional
within a Euclidean BH interpretation directly from the localized action in 
eq.~\eqref{IndependentCartanAction}.
Extremizing this constrained functional and requiring the resulting entropy
to be real and positive recovers the nonlinear charge restriction, 
further reducing the independent parameters of the known supersymmetric 
black-hole family to four.
Finally, we compare the resulting Euclidean charge locus with the Lorentzian
physical black-hole solutions of \cite{Bobev:2023bxl}, 
identifying our Euclidean entropy condition with the Lorentzian horizon 
extremality condition of these solutions.

\subsection{Supersymmetry Condition}
\label{susycond}

We proceed to derive the general spinor-weight relation that was presented in 
eq.~\eqref{IndependentCartanWeightsRelation} for the complex Euclidean 
Killing-spinor pair in eq.~\eqref{EuclideanActivePair}. 
Here we first take a gauge-covariant approach to the spinor invariance. 
We first formulate the combined isometry and gauge action on charged spinors. 
We generalize the Kosmann spinorial Lie derivative \cite{Kosmann:1971} 
to a gauge-covariant operator defined without fixing a gauge. 
This operator incorporates charged derivatives and a compensating gauge term 
expressed through moment maps \cite{Berline:1983aa}.
We then specialize to the isometry generated by the supersymmetric
Killing vector, combined with a compensating R-symmetry transformation
\cite{Ortin:2002qb}.
We first prove directly from our complex Euclidean KSEs that both Killing 
spinors are invariant under this combined action.
We highlight that our approach makes invariance under the combined
action manifest without fixing a gauge, whereas previous work evaluated
the gauge-dependent Kosmann derivative in specified gauges.
For example, in Appendix~A of \cite{Couzens:2025ghx}, the Kosmann
derivative vanishes in the chosen gauge, which should be distinguished 
from the gauge-independent invariance established here.
We then proceed to determine the spinor weights under the spacetime rotations 
and identify the constant R-symmetry parameters, combining them with 
the gauge-covariant spinor invariance to derive the relation among 
spacetime and R-symmetry weights.

\paragraph{Gauge-covariant spinor invariance.}
An isometry generated by a Killing vector $V$ preserves the gauge
background if its action on the potentials can be compensated by
a gauge transformation \cite{Ortin:2002qb}.
This compensation can be expressed through scalar moment maps
$\mu_I(V)$ satisfying \cite{Berline:1983aa}
\be
\label{BPSLiftMomentMaps}
\lambda_I(V)\equiv\mu_I(V)+\iota_VA_I,
\quad
\dd\mu_I(V)=\iota_VF_I
\qquad\implies \qquad 
\mathcal L_VA_I=\dd\lambda_I(V).
\ee
Thus the gauge transformation with parameter $-\lambda_I(V)$
cancels the change of $A_I$ under the isometry.
From eq.~\eqref{BPSLiftMomentMaps} in a Killing-invariant gauge, 
$\lambda_I(V)$ is constant, as in eq.~\eqref{BoundaryLiftConstant}.

For the opposite-charge spinor components in eq.~\eqref{EuclideanActivePair},
we write the charged derivatives appearing in eq.~\eqref{EuclideanGravitino} as
\be
\label{BPSChargedConnection}
\qquad D^\pm_\mu=\nabla_\mu \pm \mathrm{i} g(A_{1\mu}+A_{2\mu}),
\qquad\chi^+=\chi^1,\quad \chi^-=\chi^2.
\ee
The combined action on charged spinors is generated by the generalized 
gauge-covariant spinorial Lie derivative, defined as 
\begin{align}
\label{BPSTwistedLieDerivative}
\mathcal L_V^{\rm cov}\chi^\pm
&\equiv \left[V^\mu D^\pm_\mu
+\frac18(\dd V)_{\mu\nu}\gamma^{\mu\nu}
\pm \mathrm{i} g\sum_{I=1}^2\mu_I(V)\right]\chi^\pm\,
\end{align}
whereas the Kosmann derivative \cite{Kosmann:1971} reads
\be
\label{Kosmann}
\mathcal L_V^{\rm Kos}\chi^\pm
\equiv
\left[V^\mu\nabla_\mu
+\frac18(\dd V)_{\mu\nu}\gamma^{\mu\nu}\right]\chi^\pm.
\ee
For the supersymmetric Killing vector $K$ and the bilinear moment maps
specified below, we will show that
\begin{align}
\label{spinorinvariance}
\mathcal L_K^{\rm cov}\chi^\pm	
&=\mathcal L_K^{\rm Kos}\chi^\pm
\pm \mathrm{i} g\sum_{I=1}^2\lambda_I(K)\chi^\pm=0, 
\end{align}
where the first equality follows from
eqs.~\eqref{BPSLiftMomentMaps}--\eqref{Kosmann}.
The gauge-dependent Kosmann derivative in a chosen gauge then follows from 
this vanishing and the second equality of eq.~\eqref{spinorinvariance}.

Let us account first for each term in the generalized operator in 
eq.~\eqref{BPSTwistedLieDerivative}.
The first term measures the change of the spinor along the Killing
flow, using both the Levi--Civita spin and gauge connections 
to compare spinors at neighbouring points.
The second term accounts for the rotation of the local orthonormal
frame induced by the variation of the Killing vector from point to
point, and gives the corresponding action on the spinor.
Together with differentiation along the flow using only the Levi--Civita spin
connection, it gives the Kosmann derivative in eq.~\eqref{Kosmann}.
The final term provides the remaining rotation in the internal space.
Together with the gauge-connection contribution in the first term, 
it compensates the change of gauge connections under the
Killing flow, as expressed by eq.~\eqref{BPSLiftMomentMaps}, 
thus leaving them invariant under the combined action.
Adding constants to the moment maps changes this combined action
by a constant internal rotation.
This complexified operator in eq.~\eqref{BPSTwistedLieDerivative} 
then applies to the complex fields of the Euclidean theory.

For the bilinear supersymmetric Killing vector $K$, we use the moment maps 
already introduced in eqs.~\eqref{Equivariant2form}, 
\eqref{CartanEquivariantCurvatures}.
The scalar relations in eqs.~\eqref{CartanDifferentialBilinears},
together with eq.~\eqref{EuclidTransverseDilaton} under the map
in eq.~\eqref{CartanSourceSubstitution} and the contraction of
eq.~\eqref{nonequal1formrelation} with $K$, give
\be
\label{BPSSupersymmetricMoments}
\mu_I=-X_IS_3,\qquad
\dd\mu_I=\iota_KF_I,\qquad
\iota_K\dd X_I=0.
\ee
We now show that with these moment maps, the generalized operator 
$\mathcal L_K^{\rm cov}$ annihilates both Killing spinors as in 
eq.~\eqref{spinorinvariance}.
Expanding the 1-form differential relation in
eq.~\eqref{EuclideanDifferentialBilinearsUnequal} gives
\begin{align}
\label{BPSVectorDerivative}
\dd K={}&-S_3(X_1^{-1}F_1+X_2^{-1}F_2)
-\mathrm{i}X_1X_2\iota_K\star G+2g(X_1X_2)^{-2}T
-\dd\log(X_1X_2)\wedge K.
\end{align}

For the non-null gauged active spinor branch specified in
Section~\ref{BilinearRelations}, 
the Fierz identity in eq.~(A.3.3) of \cite{Levi:2026fpa},
applied with our Euclidean conventions, shows that Clifford
multiplication by $K$ and $T$ maps each spinor to a multiple
of itself, giving
\be
\label{BPSSpinorFierz}
K^\mu\gamma_\mu\chi^\pm= \mp \mathrm{i} S_3\chi^\pm,
\qquad
T_{\mu\nu}\gamma^{\mu\nu}\chi^\pm= \pm 6\mathrm{i} S_3\chi^\pm,
\ee
where the coefficients are fixed by contracting with
$\overline{\chi^\mp}$ and using eq.~\eqref{Killingnorm}
and the trace of eq.~\eqref{EuclideanTwoFormEndomorphism},
respectively.
We also note that the $G$-contraction arising from the differential KSE in
eq.~\eqref{EuclideanGravitino} can be rewritten using
the Clifford dual in eq.~\eqref{EuclideanCliffordDual} as
\be
\label{BPSFourFormDuality}
K^\mu\gamma_\mu{}^{\nu\rho\lambda\tau}G_{\nu\rho\lambda\tau}
=-12\mathrm{i}(\iota_K\star G)_{\alpha\beta}\gamma^{\alpha\beta}.
\ee

Let $\mathcal A_\pm$ denote the operator of the algebraic KSE 
acting on $\chi^\pm$ in eq.~\eqref{EuclideanAlgebraicSym}.
Then substituting the differential KSE in eq.~\eqref{EuclideanGravitino}
and eqs.~\eqref{BPSSupersymmetricMoments}--\eqref{BPSFourFormDuality}
into eq.~\eqref{BPSTwistedLieDerivative} yields
\begin{align}
\label{BPSLiftCancellation}
\mathcal L_K^{\rm cov} \chi^\pm
&=(K^\mu\gamma_\mu \mp \mathrm{i} S_3)
\left[\mathcal A_\pm
+\frac58\gamma^\nu\partial_\nu\log(X_1X_2)\right]\chi^\pm
=0,
\end{align}
where the first term in square brackets vanishes by 
eq.~\eqref{EuclideanAlgebraicSym}, and for the second term in these 
brackets, anticommuting $K^\mu\gamma_\mu$ through the scalar-gradient term makes 
that term vanish by 
eqs.~\eqref{BPSSupersymmetricMoments}, \eqref{BPSSpinorFierz}.
This completes the proof of invariance of both spinors under the combined
spacetime and R-symmetry action independently of any specific gauge choice.

\paragraph{Spacetime and R-symmetry rotational weights.}
On the physical spacetime $\mathbb R^2\times S^5$, the four $2\pi$-periodic 
generators in eq.~\eqref{KillingVectorExtension} rotate the $\mathbb R^2$ plane 
and the three complex coordinates of $S^5\subset\mathbb C^3$.
The commuting generators have commuting Kosmann derivatives \cite{Ortin:2002qb}:
\be 
\label{commutingKosmann}
[\mathcal L_{V_a}^{\rm Kos},
\mathcal L_{V_b}^{\rm Kos}]
=\mathcal L_{[V_a,V_b]}^{\rm Kos}=0.
\ee
We assume that each of these four rotations preserves the metric,
the scalars $X_I$, and the field strengths $F_I$ and $G$.
We apply separately to the two Cartan fields the arguments detailed
in Section~\ref{filling} for the existence of a global auxiliary
potential and the extraction of its global torus-invariant part. 
Since $H^2(\mathbb R^2\times S^5)=0$, both Cartan field strengths admit
global potentials, and torus averaging simultaneously extracts their 
gauge-equivalent global torus-invariant parts $\hat A_I$.
Let $\mathcal K^+$ denote the space of Killing-spinor solutions
for the $\chi^+$ component on the fixed bosonic background,
satisfying the KSEs in eqs.~\eqref{EuclideanGravitino},
\eqref{EuclideanAlgebraicKSE}.
Since each rotation preserves the background, including the potentials $\hat A_I$
in the chosen gauge, covariance of the KSEs implies that it maps any Killing 
spinor to another solution of the same equations \cite{Ortin:2002qb}.
Then the Kosmann derivatives along $V_a$ preserve $\mathcal K^+$:
\begin{align}
\label{BPSRotationalKSEAction}
\psi\in\mathcal K^+
&\quad\Longrightarrow\quad
\mathcal L_{V_a}^{\rm Kos}\psi\in\mathcal K^+.
\end{align}

Each generator has a fixed set $Z_a$ of codimension two, with unit
rotation on its normal plane.
At a fixed point $p_a\in Z_a$, we choose an oriented orthonormal
normal frame $(x_a,y_a)$.
Since $V_a(p_a)=0$, only the $\dd V_a$ term in the Kosmann
derivative in eq.~\eqref{Kosmann} contributes on the fixed set.
The unit normal rotation then gives $(\dd V_a)_{x_ay_a}(p_a) = 2$, 
and for any smooth spinor $\psi$ \cite{BenettiGenolini:2024lbj},
\be
\label{BPSFixedCircleSpinAction}
\mathcal L_{V_a}^{\rm Kos}\psi (p_a)
=\frac12\gamma^{x_ay_a}\psi(p_a),
\ee
where for any orthonormal pair at any point, the Euclidean Clifford
algebra in eq.~\eqref{EuclideanClifford} gives
\be
\label{squared2Cliff}
(\gamma^{x_ay_a})^2=-\mathds{1}.
\ee
Along any smooth curve, the differential KSE in eq.~\eqref{EuclideanGravitino} 
becomes a homogeneous linear ordinary differential equation for the spinor.
Uniqueness of its initial-value problem implies that two Killing-spinor 
solutions agreeing at one point agree along every curve through that point and 
hence throughout the connected spacetime \cite{Ferrero:2021etw}.
Applying eq.~\eqref{BPSFixedCircleSpinAction} twice and using
eq.~\eqref{squared2Cliff} makes the squared Kosmann derivative
proportional to the original Killing spinor at $p_a$.
Both are KSE solutions by eq.~\eqref{BPSRotationalKSEAction},
so uniqueness extends this proportionality throughout spacetime:
\be
\label{BPSRotationalWeightSquare}
(\mathcal L_{V_a}^{\rm Kos})^2
=-\frac14\mathds{1}
\quad\text{on }\mathcal K^+.
\ee
This splits $\mathcal K^+$ into the $\pm \mathrm{i}/2$ eigenspaces of each Kosmann 
derivative, 
which is annihilated by a quadratic polynomial with two distinct roots.
Since the Kosmann derivatives commute by eq.~\eqref{commutingKosmann}, 
the solution space decomposes into their common eigenspaces:
\be
\label{BPSJointWeightSpaces}
\mathcal K^+=\bigoplus_{\vec{s}}\mathcal K^+_{\vec{s}},
\qquad
\vec{s}\equiv(s_0,s_1,s_2,s_3)\in\{\pm1\}^4,
\ee
where the four signs vary independently, and
the preserved spinor decomposes into components
in these common eigenspaces:
\be
\label{Killingeigencomponents}
\chi^+=\sum_{\vec{s}}\chi^+_{\vec{s}},
\qquad \chi^+_{\vec{s}}\in\mathcal K^+_{\vec{s}},
\ee
which satisfy
\be
\label{BPSTorusSpinorWeights}
\mathcal L_{V_a}^{\rm Kos}\chi^+_{\vec{s}}
=\frac{\mathrm{i}}{2}s_a\chi^+_{\vec{s}},
\qquad s_a=\pm1, \quad a=0,1,2,3.
\ee

For $K$ in eq.~\eqref{KillingVectorExtension}, we denote by
$\hat\lambda_I$ the constant values of $\lambda_I(K)$ in
eq.~\eqref{BPSLiftMomentMaps} evaluated in this invariant gauge.
The expression for $\mathcal L_K^{\rm cov}$ in eq.~\eqref{spinorinvariance} 
then acts diagonally on the components in eq.~\eqref{Killingeigencomponents}.
The invariance established in eq.~\eqref{BPSLiftCancellation}
therefore implies, for each nonzero spinor component,
\be
\label{BPSSignedWeightConstraint}
\sum_{a=0}^3s_a\varepsilon_a
+2g\sum_{I=1}^2\hat\lambda_I=0.
\ee
We consider a nonzero Killing-spinor component with $s_a=+1$ for all four 
generators in the chosen orientations.
If we identify the constants $2g\hat\lambda_I$ as the R-symmetry rotational
weights, $b_I$, then eq.~\eqref{BPSSignedWeightConstraint} gives 
eq.~\eqref{IndependentCartanWeightsRelation} on the physical spacetime.
For the extension of Section~\ref{FullCartanExtensions} on the filling,
the identification in eq.~\eqref{independentmoments} also
recovers eq.~\eqref{IndependentCartanWeights}.
For the black-hole interpretation, the dictionary in
eq.~\eqref{IndependentWeightDictionary} then gives the
chemical-potential constraint in
eq.~\eqref{IndependentChemicalConstraint}.

\subsection{Entropy Condition}
\label{RealMaxEntropy}

The localized action in eq.~\eqref{CartanDirectLocalizedAction} describes
complex supersymmetric Euclidean solutions. 
We now invoke the interpretation of this action within Euclidean black-hole
thermodynamics in eq.~\eqref{IndependentCartanAction}, using the quantum 
statistical relation as in \cite{Bobev:2023bxl} to construct the  
entropy functional.
Extremizing this constrained functional
and requiring the result to be real and positive
then yields the additional nonlinear charge relation.
This relation remains entirely within the Euclidean theory, with the 
chemical-potential constraint in eq.~\eqref{IndependentChemicalConstraint} 
from the supersymmetry condition maintained throughout. 
The former constraint discussed in Section~\ref{susycond} reduces the 
six-parameter solution family to five parameters. 
The entropy condition here then removes one further independent parameter and, 
on the physical branch, reproduces the physical charge relation satisfied by 
regular supersymmetric black holes.

The supersymmetric Euclidean solutions of \cite{Bobev:2023bxl} are
themselves regular complex ones. 
Their analytic continuation to Lorentzian signature, however, generically
has complex fields or causal pathologies such as closed timelike curves.
As noted the additional entropy condition concerns their real black-hole 
interpretation and is derived from Euclidean thermodynamics. We subsequently 
compare it with the extremal-horizon condition on the Lorentzian solution. 
The latter provides a geometric interpretation of the resulting charge
restriction and identifies the physical solutions and parameter domain.
The constrained entropy extremization was proposed for rotating
AdS$_5$ and AdS$_7$ black holes in \cite{Hosseini:2017mds,Hosseini:2018dob}. 
Its derivation from complex supersymmetric Euclidean solutions, including the 
nonlinear charge relation obtained by imposing entropy reality, was developed in
\cite{Cabo-Bizet:2018ehj,Cassani:2019mms}.

\paragraph{Constraining the entropy.}
We use the electric charges and angular momenta of \cite{Bobev:2023bxl}, 
with the coupling map in eq.~\eqref{BlackHoleCouplingMap}, and define
\be
\label{EntropyChargeNormalization}
q_I\equiv\frac{Q_I}{2\mathfrak g},
\qquad
\kappa\equiv\frac{2N^3}{3}
=\frac{\pi^2}{8G_N\mathfrak g^5},
\ee
where the normalization follows from eqs.~\eqref{IndependentWeightDictionary}
and \eqref{LocalizedEqualChargeAction}. 
In particular, using eq.~\eqref{IndependentWeightDictionary} and the
supersymmetric energy relation in eq.~(4.31) of \cite{Bobev:2023bxl},
the quantum statistical relation becomes
\be
\label{SupersymmetricQuantumStatisticalRelation}
I_{\rm BH}=-S+\sum_{I=1}^{2}q_I\Delta_I
-\sum_{i=1}^{3}J_i\omega_i.
\ee
Introducing a Lagrange multiplier for the weight constraint in 
eq.~\eqref{IndependentChemicalConstraint}, we then extremize
\begin{align}
\label{ConstrainedEntropyFunctional}
\mathcal S(\Delta_I,\omega_i,\Lambda)
={}&-I_{\rm BH}+\sum_{I=1}^{2}q_I\Delta_I
-\sum_{i=1}^{3}J_i\omega_i
-\Lambda\left(\sum_{I=1}^{2}\Delta_I
-\sum_{i=1}^{3}\omega_i-2\pi \mathrm{i}\right),
\end{align}
where the potentials and multiplier are complex. 

Varying the entropy functional with respect to $\Delta_I$ and $\omega_i$
independently, while holding the charges fixed, gives the stationarity
conditions
\begin{subequations}
\label{EntropyStationarity}
\begin{align}
\label{electric}	
2I_{\rm BH}&=(q_I-\Lambda)\Delta_I,
\\*
\label{angular}
I_{\rm BH}&=(J_i-\Lambda)\omega_i.
\end{align}
\end{subequations}
Substituting these stationarity conditions back into the entropy functional in
eq.~\eqref{ConstrainedEntropyFunctional} gives the extremized entropy,
\be
\label{EntropyMultiplier}
\mathcal S_*=2\pi \mathrm{i}\Lambda.
\ee
Requiring the extremized entropy in eq.~\eqref{EntropyMultiplier}
to be real and positive amounts to
\begin{subequations}
\label{EntropyRealityRoot}
\begin{align}
\Lambda &\equiv -\mathrm{i}s,
\\*
\label{positiveentropy}
s &\equiv\frac{S_{\rm BH}}{2\pi}>0.
\end{align}
\end{subequations}

Squaring and multiplying the two electric relations in eq.~\eqref{electric} 
gives 
\be
\label{topolynomial}
\prod_{I=1}^{2}(\Lambda-q_I)^2
=\frac{16I_{\rm BH}^4}{\Delta_1^2\Delta_2^2}
=\frac{\kappa I_{\rm BH}^3}{\omega_1\omega_2\omega_3}
=\kappa\prod_{i=1}^{3}(J_i-\Lambda),
\ee
where the second equality uses the localized action in 
eq.~\eqref{IndependentCartanAction} with the normalization in
eq.~\eqref{EntropyChargeNormalization}, and the last one 
uses the angular relations in eq.~\eqref{angular}.
Rearranging eq.~\eqref{topolynomial} gives the quartic equation for the entropy, 
as in \cite{Choi:2018hmj}, expressed here in terms of the multiplier, 
\be
\label{EntropyQuartic}
P(\Lambda)\equiv \prod_{I=1}^{2}(\Lambda-q_I)^2
+\kappa\prod_{i=1}^{3}(\Lambda-J_i)
=\Lambda^4+p_3\Lambda^3+p_2\Lambda^2+p_1\Lambda+p_0=0,
\ee
where
\begin{subequations}
\label{EntropyQuarticCoefficients}
\begin{align}
p_3&=\kappa-2(q_1+q_2),\\*
p_2&=q_1^2+q_2^2+4q_1q_2-\kappa(J_1+J_2+J_3),\\*
p_1&=\kappa(J_1J_2+J_2J_3+J_3J_1)-2q_1q_2(q_1+q_2),\\*
p_0&=q_1^2q_2^2-\kappa J_1J_2J_3.
\end{align}
\end{subequations}

For real conserved charges these coefficients are real. 
The real and imaginary parts of eq.~\eqref{EntropyQuartic} then give, 
in terms of $s$ defined in eqs.~\eqref{EntropyRealityRoot},
\begin{subequations}
\label{EntropyRealityEquations}
\begin{align}
\label{even}	
s^4-p_2s^2+p_0&=0,
\\
\label{odd}
p_3s^2-p_1&=0.
\end{align}
\end{subequations}
On the generic locus where $p_3$ is nonzero, substituting in $s^2=p_1/p_3$ 
yields the nonlinear charge relation
\be
\label{EntropyChargeConstraint}
p_1^2-p_2p_1p_3+p_0p_3^2=0.
\ee
This is an additional real relation among the five conserved charges
remaining after imposing the supersymmetric condition. 
It therefore reduces the parameters of the solution family 
to four independent parameters.

\paragraph{Relating Euclidean entropy and Lorentzian extremality.}
Comparison of the last relation with the nonlinear charge relation in eq.~(4.39) 
of \cite{Bobev:2023bxl} identifies the known regular Lorentzian 
spherical-horizon black holes with the minus-sign root of the discriminant for 
$p_1/p_3$ obtained from eqs.~\eqref{EntropyRealityEquations}, 
\eqref{EntropyChargeConstraint}, giving 
\be
\label{EntropyPhysicalBranch}
\frac{p_1}{p_3}
=\frac{p_2-\sqrt{p_2^2-4p_0}}{2}>0,
\ee
with the square root real and nonnegative. 
Indeed, expressing eq.~\eqref{EntropyPhysicalBranch} in terms of the
charges using eqs.~\eqref{EntropyChargeNormalization} and
\eqref{EntropyQuarticCoefficients} reproduces eq.~(4.39) of
\cite{Bobev:2023bxl}.
Taking the positive square root of $s^2$ in 
eq.~\eqref{odd}, as required by
eq.~\eqref{positiveentropy}, then gives the BH entropy
\begin{subequations}
\label{RealBlackHoleEntropy}
\begin{align}
S_{\rm BH}&=2\pi\sqrt{\frac{p_1}{p_3}}
=2\pi\sqrt{
\frac{2q_1q_2(q_1+q_2)-\kappa(J_1J_2+J_2J_3+J_3J_1)}
{2(q_1+q_2)-\kappa}}\\*
\label{special}
&=2\pi\sqrt{\frac{p_2-\sqrt{p_2^2-4p_0}}{2}}
\end{align}
\end{subequations}
in agreement with eq.~(4.40) of \cite{Bobev:2023bxl} and the extremization 
result in eq.~(5.9) of \cite{Nahmgoong:2019hko}.
At the special locus where $p_3$ vanishes, eq.~\eqref{odd} requires
$p_1$ to vanish. The BH entropy is then given directly by eq.~\eqref{special}, 
still corresponding to the smaller positive root for $s^2$ in eq.~\eqref{even}.

The charge relation and entropy in eqs.~\eqref{EntropyPhysicalBranch} and 
\eqref{RealBlackHoleEntropy}, respectively, complete the connection between the 
Euclidean localized action in eq.~\eqref{IndependentCartanAction} and the known 
physical four-parameter black-hole family from \cite{Bobev:2023bxl}. 
Let us further elucidate the relation between the Euclidean entropy conditions 
in eqs.~\eqref{EntropyStationarity}--\eqref{EntropyRealityRoot}, 
and the Lorentzian geometry of the explicit BH solution. 
Within the supersymmetric Lorentzian branch of \cite{Bobev:2023bxl}, 
the horizon and extremality conditions in eqs.~(4.2) and~(4.35)
fix the mass parameter to eq.~(4.36), with the extremal outer horizon radius 
given by eq.~(4.37) therein.
Expressing this mass restriction in terms of the conserved charges gives the 
nonlinear charge constraint in eq.~(4.39)  of \cite{Bobev:2023bxl}, 
already found to agree with our eqs.~\eqref{EntropyPhysicalBranch}, 
\eqref{EntropyChargeConstraint}.
Our Euclidean entropy condition therefore corresponds to
Lorentzian horizon extremality on this branch.

\section{Conclusions}
\label{LaFin}

In this work we have established relative equivariant localization
in seven-dimensional maximal Cartan-gauged supergravity, 
with both Abelian gauge fields and the massive 3-form gauge field active. 
Localization at the single fixed point of an auxiliary 8-manifold filling 
reproduces the general two-charge, three-angular-momentum supersymmetric 
AdS$_7$ black-hole action of \cite{Bobev:2023bxl}.
This is the first seven-dimensional localization with the higher-form gauge 
field active, and the first application of relative equivariant localization 
in an odd spacetime dimension beyond $d=5,11$, demonstrated in the original 
proposal of \cite{BenettiGenolini:2026cdw}.

We start by first providing a Euclidean formulation of the seven-dimensional 
maximal Cartan-gauged theory with its full bosonic field content, the action, 
and complete KSEs in terms of the field strengths. 
This provides the Euclidean complement to the Lorentzian field-strength 
formulation established in \cite{Levi:2026fpa}.
Seven dimensions are unique among $d\leq11$ in that passing from the Lorentzian 
to Euclidean Clifford algebra unavoidably changes the spinor reality conditions 
from symplectic-Majorana to Majorana.
Consequently, the standard spinor-bilinear construction based on
charge-conjugate Killing-spinor pairs cannot be carried over from 
the Lorentzian theory, requiring a new bilinear pairing adapted
to the seven-dimensional setting.
Our construction instead relies crucially on the internal spinor branch 
$\Ker\Gamma_-$ of the gauged active class identified in \cite{Levi:2026fpa}, 
using two independent, oppositely R-charged Killing spinors 
on the same complex background.
We thereby first introduce a set of seven-dimensional Euclidean mixed bilinears, 
formed by pairing one of these independent Killing spinors with the Majorana 
conjugate of the other.
Their bilinear analysis uses Majorana-conjugate KSEs, which maintain
the original complex background fields. 
We provide here the necessary Euclidean bilinear analysis in detail, 
which complements the Lorentzian one in \cite{Levi:2026fpa}.

Moreover, we show here that the extension to independent Cartan charges in the 
maximal Cartan-gauged theory preserves much of the equal-charge case, 
building on \cite{Levi:2026fpa}. 
Using a geometric-mean scalar and dressed symmetric and antisymmetric fluxes 
isolates the additional supersymmetric input into one algebraic gauging 
correction and two new algebraic bilinear projections. 
These suffice to extend the bilinear relations and the relative equivariant 
localization. The underlying reorganization of the Cartan KSEs also applies 
in Lorentzian signature, providing a convenient route from the equal-charge 
foundation in \cite{Levi:2026fpa} to applications in the maximal 
Cartan-gauged theory in the corresponding active class. 
Together, the action, KSEs and bilinear analysis therefore support 
broader applications of the seven-dimensional Cartan-gauged theory.

Building on \cite{BenettiGenolini:2026cdw}, we construct from these data 
a global, gauge-invariant relative equivariant cohomology class, 
without an explicit solution metric or local representatives 
for the gauge potentials. 
The weighted 4-form closure of the transverse geometry, 
which corresponds to the exact warp-dependent torsion 1-form
$\mathcal W_4$ in \cite{Levi:2026fpa}, 
controls the cancellation of residual gauging terms 
required for relative equivariant closure.
The self-duality relation of the massive 3-form gauge field fixes
the quartic anomaly class, uncovering the seven-dimensional mechanism 
behind the quartic structure obtained in \cite{BenettiGenolini:2026cdw} 
through integration over $S^4$.
Localization of this anomaly class thus yields the quadratic dependence 
of the black-hole action on each of the two R-symmetry chemical potentials. 
At the fixed point of the auxiliary filling, regularity determines the scalar
component of the extended equivariant 4-form from the two gauge-field moment
maps, leaving no additional independent data from the massive 3-form.

We have also proved directly from the complex Euclidean KSEs that 
the generic Killing spinor in this active class is invariant 
under the spacetime isometry generated by the supersymmetric
Killing vector, combined with the R-symmetry action, formulated through 
a generalized gauge-covariant spinorial Lie derivative.
This gauge-covariant invariance determines the ordinary 
gauge-dependent Kosmann derivative in any chosen gauge.
We then combine the rotational spinor weights  on physical spacetime 
with this gauge-covariant invariance to explicitly derive the general 
spinor-weight relation.
This yields the chemical-potential constraint which restricts the general
six-parameter black-hole family in \cite{Bobev:2023bxl}.

Finally, extremization of the constrained entropy functional
constructed from the localized Euclidean action,
together with reality and positivity of the entropy, yields the
additional nonlinear charge relation and the physical black-hole entropy. 
On the known spherical-horizon branch, this Euclidean restriction is equivalent 
to Lorentzian horizon extremality and selects the charge locus 
of the regular four-parameter black-hole family in \cite{Bobev:2023bxl}.
Thus both the supersymmetry constraint and the additional thermodynamic
restriction follow within the Euclidean analysis, with their distinct
meanings sharpened.

Natural next steps include higher-derivative corrections, guided by
the six-dimensional SCFT anomaly data in 
\cite{Bobev:2015kza,Nahmgoong:2019hko,Ohmori:2021dzb}. 
Other seven-dimensional spacetime topologies, including those
of the black-hole solutions constructed from $L^{p,q,r}$ geometries
in \cite{Bobev:2025xan},
motivate a systematic treatment of admissible fillings 
and an investigation of a Sasakian-volume interpretation of the action. 
Additional directions include extending seven-dimensional localization
to other solution classes identified in \cite{Levi:2026fpa},
further building on the framework developed there.
All of these are left to future work, and we plan to report on it soon.

\begin{acknowledgments}
We are greatly indebted to James Sparks for his sustained interest 
in this project, for comments on an earlier draft of the manuscript 
and for his invaluable encouragement and support.
FG thanks him for many insightful discussions 
and for clarifying several mathematical points. 
FG is supported by an STFC studentship.	
ML has been supported by the Science and Technology Facilities Council 
(STFC) Rutherford Grant ST/V003895 \textit{``Harnessing QFT for 
Gravity''}, and by the Mathematical Institute, University of Oxford.	
\end{acknowledgments}

\appendix

\section{Appendices}

\counterwithin{equation}{subsection}

\subsection{Euclidean Clifford Algebra in Dimension 7}
\label{7ECliff}

We present the Euclidean Clifford algebra Cliff$(7)$ and conventions 
following the Lorentzian ones in Appendix~A.1 of \cite{Levi:2026fpa}. 
Tangent indices are $a=0,\ldots,6$, with $0$ denoting the continued time 
direction.
The coframe continuation used in Section~\ref{euclidtheory} preserves
the Clifford-valued 1-form when
\begin{subequations}
\label{EuclideanGammaMap}
\begin{align}
\gamma_E^0&= \mathrm{i} \gamma_L^0= - \mathrm{i} \gamma_{L,0},
\\*
\gamma_E^i&=\gamma_L^i,\qquad i=1,\ldots,6.
\end{align}
\end{subequations}
Indeed, $\gamma_{L,0}e_L^0\mapsto\gamma_{E,0}e_E^0$.
It follows from eqs.~(A.1.1)--(A.1.2) of \cite{Levi:2026fpa} that
\be
\label{EuclideanClifford}
\{\gamma_E^a,\gamma_E^b\}=2\delta^{ab}\mathds{1}_8,
\qquad
(\gamma_E^a)^\dagger=\gamma_E^a.
\ee
The irreducible Clifford
representation is fixed by continuing
$\gamma_{L,0123456}=-\mathds{1}_8$ in
eq.~(A.1.7) of \cite{Levi:2026fpa}, which gives
\be
\label{EuclideanCliffordVolume}
\gamma_{E,0123456}= \mathrm{i} \mathds{1}_8.
\ee
Henceforth we omit the subscript $E$ on the gamma matrices.

The charge-conjugation matrix $C$ is unchanged: it is symmetric and
unitary, and
\begin{subequations}
\label{EuclideanConjugationMatrices}
\begin{align}
\label{EuclideanCTranspose}
(\gamma^a)^T&=-C\gamma^aC^{-1},
& C^T&=C,
\\*
\label{EuclideanBConjugation}
(\gamma^a)^*&=-B_E\gamma^aB_E^{-1},
& B_E&=C,
\end{align}
\end{subequations}
where the complex-conjugation matrix $B$ is modified, and combining the 
unitarity of $C$ with these equations yields
\be 
\label{7EuclidMajoran}
B_E^*B_E=\mathds{1}_8\, .
\ee
The representation $C=\mathds{1}_8$ adopted in eq.~(A.1.5) of 
\cite{Levi:2026fpa} remains valid. 
The Euclidean gamma matrices are then purely imaginary and antisymmetric.
The Majorana and Euclidean charge conjugates are
\begin{subequations}
\label{EuclideanChargeConjugates}
\begin{align}
\overline\epsilon&=\epsilon^T C,\\*
\label{EuclideanChargeDefinition}
\epsilon^{c_E}&=C^{-1}\epsilon^*,
\end{align}
\end{subequations}
so that
\be
\label{EuclidMajoranaChargeConjugate}
\overline{\epsilon^{c_E}}=\epsilon^\dagger,
\ee
and 
\be
\label{EuclideanChargeSquare}
(\epsilon^{c_E})^{c_E}=\epsilon,
\ee
so that the Euclidean Clifford algebra admits an ordinary Majorana condition
which differs from the Lorentzian case, for which
$B_L= - \mathrm{i} C\gamma_L^0$ and $B_L^*B_L=-\mathds{1}_8$.

\subsection{Conventions for 7D Euclidean Bilinears}
\label{EuclideanBilinearConventions}

We present here the conventions of \cite{Levi:2026fpa} adapted to the
Euclidean Clifford algebra in Appendix~\ref{7ECliff}. 
The Euclidean counterpart of the spacetime orientation in eq.~(A.3.7) 
of \cite{Levi:2026fpa} is given by
\be
\label{EuclideanOrientation}
\epsilon_{0123456}=\epsilon^{0123456}=+1.
\ee
With eq.~\eqref{EuclideanOrientation}, the complex-linear Hodge star
retains the index ordering of eq.~(A.3.10) of \cite{Levi:2026fpa}, 
and from the convention-independent Hodge square in eq.~(A.14.2) 
of \cite{Levi:2026fpa} we have
\be
\label{EuclideanHodgesquare}
\star^2\alpha
=\alpha.
\ee

From the irreducible Clifford representation 
in eq.~\eqref{EuclideanCliffordVolume} and the orientation in 
eq.~\eqref{EuclideanOrientation}, we get the Clifford dual
\be
\label{EuclideanCliffordDual}
\gamma_{a_1\ldots a_r}
= - \mathrm{i} \,\frac{(-1)^{\lfloor r/2\rfloor}}{(7-r)!}
\epsilon_{a_1\ldots a_r b_1\ldots b_{7-r}}
\gamma^{b_1\ldots b_{7-r}},
\qquad 0\leq r\leq7,
\ee
as the Euclidean counterpart of eq.~(A.3.8) of \cite{Levi:2026fpa}.

For commuting spinors, the Majorana flip and Fierz identities in
eqs.~(A.3.1), (A.3.3) of \cite{Levi:2026fpa} are unchanged, with
Euclidean contractions. The complex-conjugation formula in
eq.~(A.3.2) of \cite{Levi:2026fpa}, however, becomes
\be
\label{EuclideanBilinearConjugation}
(\overline\lambda M\chi)^*
=\overline{\lambda^{c_E}}\,M^{c_E}\chi^{c_E},
\qquad
M^{c_E}=C^{-1}M^*C.
\ee
The overall plus sign in eq.~\eqref{EuclideanBilinearConjugation} follows
from eqs.~\eqref{EuclideanConjugationMatrices}
and the spinor definitions in eqs.~\eqref{EuclideanChargeConjugates}.

\subsection{Conjugates of Euclidean Killing Spinors}
\label{EuclideanConjugateKSEs}

We derive the Euclidean counterparts of eqs.~(A.9.1)--(A.9.3) of
\cite{Levi:2026fpa} from the Euclidean KSEs 
in eqs.~\eqref{EuclideanGravitino} and \eqref{EuclideanAlgebraicKSE}. 
The spacetime conjugates are defined in eqs.~\eqref{EuclideanChargeConjugates}, 
while the internal matrices $\Gamma_{\hat I}$ and $\Gamma_5$ are real and 
symmetric in the basis of eq.~(2.30) of \cite{Levi:2026fpa}. 

As the Euclidean bosonic geometry is in general complex, we denote the curved 
Clifford matrices and spin connection constructed from the conjugate coframe by
\begin{subequations}
\label{EuclideanConjugateGeometry}
\begin{align}
\label{EuclideanConjugateConnection}
\widehat\nabla_\mu&\equiv\partial_\mu
+\frac14(\omega_\mu{}^{ab})^*\gamma_{ab}, \\*	
\label{EuclideanConjugateCurvedClifford}
\widehat\gamma_\mu&\equiv(e_\mu{}^a)^*\gamma_a,
\qquad \qquad \widehat\gamma^\mu\equiv(e_a{}^\mu)^*\gamma^a.
\end{align}
\end{subequations}
For a real Euclidean coframe, the hats may be omitted, whereas the bosonic
matter fields remain complex. From eqs.~\eqref{EuclideanConjugationMatrices}
we have
\be
\big(\nabla_\mu\epsilon\big)^{c_E}
=\widehat\nabla_\mu\epsilon^{c_E}, \qquad
\label{EuclideanConjugateCovariance}
\big(\gamma_\mu\big)^{c_E}=-\widehat\gamma_\mu.
\ee

Applying eqs.~\eqref{EuclideanConjugationMatrices} and
eq.~\eqref{EuclideanConjugateCovariance} to
eq.~\eqref{EuclideanGravitino}, the differential equation for the
Euclidean charge conjugate is
\begin{align}
\label{EuclideanGravitinoCharge}
\widehat\nabla_\mu\epsilon^{c_E}
=\bigg[& \mathrm{i} g\big(A_{1\mu}^*\Gamma_{\hat1}
+A_{2\mu}^*\Gamma_{\hat2}\big)
-\frac{\mathrm{i}}{4}\widehat\gamma^\nu
\big((X_1^*)^{-1}F_{1\mu\nu}^*\Gamma_{\hat1}
+(X_2^*)^{-1}F_{2\mu\nu}^*\Gamma_{\hat2}\big)
\nn\\*
&+\frac g2(X_1^*X_2^*)^{-2}\widehat\gamma_\mu
-\frac14\widehat\gamma_\mu\widehat\gamma^\nu
\partial_\nu\log(X_1^*X_2^*)
-\frac{X_1^*X_2^*}{96}
\widehat\gamma_\mu{}^{\nu\lambda\sigma\rho}
G_{\nu\lambda\sigma\rho}^*\Gamma_5\bigg]\epsilon^{c_E}.
\end{align}
Similarly, eqs.~\eqref{EuclideanAlgebraicKSE} give
\begin{subequations}
\label{EuclideanAlgebraicCharge}
\begin{align}
\label{EuclideanAlgebraicSymCharge}
0=\bigg[&\frac g2\big(X_1^*+X_2^*-2(X_1^*X_2^*)^{-2}\big)
+\frac58\widehat\gamma^\mu\partial_\mu\log(X_1^*X_2^*)
\nn\\*
&+\frac{\mathrm{i}}{16}\widehat\gamma^{\mu\nu}
\big((X_1^*)^{-1}F_{1\mu\nu}^*\Gamma_{\hat1}
+(X_2^*)^{-1}F_{2\mu\nu}^*\Gamma_{\hat2}\big)
+\frac{X_1^*X_2^*}{96}\widehat\gamma^{\mu\nu\lambda\sigma}
G_{\mu\nu\lambda\sigma}^*\Gamma_5\bigg]\epsilon^{c_E},\\*
\label{EuclideanAlgebraicAntisymCharge}
0=\bigg[&\frac g2(X_1^*-X_2^*)
+\frac18\widehat\gamma^\mu\partial_\mu\log(X_1^*/X_2^*)
\nn\\*
&+\frac{\mathrm{i}}{16}\widehat\gamma^{\mu\nu}
\big((X_1^*)^{-1}F_{1\mu\nu}^*\Gamma_{\hat1}
-(X_2^*)^{-1}F_{2\mu\nu}^*\Gamma_{\hat2}\big)\bigg]\epsilon^{c_E}.
\end{align}
\end{subequations}
Compared with eqs.~(A.9.1)--(A.9.2) of \cite{Levi:2026fpa},
the odd-degree spacetime Clifford terms acquire an additional minus sign, 
and the bosonic coefficients are complex conjugated.

For the Majorana conjugate, transposing eq.~\eqref{EuclideanGravitino}
and using eqs.~\eqref{EuclideanConjugationMatrices} 
gives an equation of the same form as eq.~(A.9.3) in \cite{Levi:2026fpa}. 
Transposing the algebraic KSEs likewise gives
\begin{subequations}
\label{EuclideanAlgebraicMajorana}
\begin{align}
\label{EuclideanAlgebraicSymMajorana}
0=\overline\epsilon\bigg[&\frac g2\big(X_1+X_2-2(X_1X_2)^{-2}\big)
+\frac58\gamma^\mu\partial_\mu\log(X_1X_2)
\nn\\*
&+\frac{\mathrm{i}}{16}\gamma^{\mu\nu}
\big(X_1^{-1}F_{1\mu\nu}\Gamma_{\hat1}
+X_2^{-1}F_{2\mu\nu}\Gamma_{\hat2}\big)
+\frac{X_1X_2}{96}\gamma^{\mu\nu\lambda\sigma}
G_{\mu\nu\lambda\sigma}\Gamma_5\bigg],\\*
\label{EuclideanAlgebraicAntisymMajorana}
0=\overline\epsilon\bigg[&\frac g2(X_1-X_2)
+\frac18\gamma^\mu\partial_\mu\log(X_1/X_2)
\nn\\*
&+\frac{\mathrm{i}}{16}\gamma^{\mu\nu}
\big(X_1^{-1}F_{1\mu\nu}\Gamma_{\hat1}
-X_2^{-1}F_{2\mu\nu}\Gamma_{\hat2}\big)\bigg].
\end{align}
\end{subequations}
The equations for $\overline{\epsilon^{c_E}}=\epsilon^\dagger$ in
eq.~\eqref{EuclidMajoranaChargeConjugate} follow by applying the same
Majorana transpose to eqs.~\eqref{EuclideanGravitinoCharge} and
\eqref{EuclideanAlgebraicCharge}.

\subsection{Bilinear Relations in Equal-Charge Sector}
\label{EuclideanActiveIdentities}

We derive the relations used in Section~\ref{BilinearAnalysis} with the
Euclidean conventions of Appendices~\ref{7ECliff} and
\ref{EuclideanBilinearConventions}. 
Throughout this appendix we assume the gauged active class of 
\cite{Levi:2026fpa}, and thus the pair of opposite-charge spinors 
in eq.~\eqref{EuclideanActivePair} satisfies 
eqs.~\eqref{EuclideanChargedScalarsVanish}, \eqref{nonzeroscalar}.

\paragraph{Geometric relations and Hodge duals.}
Applying the Fierz identity in eq.~(A.3.3) of \cite{Levi:2026fpa}
with Euclidean contractions gives eqs.~\eqref{EuclideanActiveFierz} and
\be
\label{EuclideanTwoFormEndomorphism}
T_\mu{}^\rho T_{\rho\nu}
=-S_3^2g_{\mu\nu}-K_\mu K_\nu.
\ee
This is the Euclidean counterpart of the relation obtained
from eqs.~(3.43), (A.5.5) of \cite{Levi:2026fpa} 
involving the almost complex structure,
using $S_3^2=h^2$ in the gauged active class.

The higher-rank bilinears needed for the algebraic relations in
eqs.~\eqref{EuclideanAlgebraicProjections} involve the
Euclidean counterpart of $U_3=u^{12}$ in eq.~(3.45b) of
\cite{Levi:2026fpa}.
In the gauged active class with eq.~\eqref{EuclideanChargedScalarsVanish} 
so that $s=0$, this reduces to $U_3=-S_3^{-1}K\wedge T$ from eq.~(3.49), 
using $S_3=-hc$ and $c^2=1$.
The Euclidean Fierz and Clifford-duality formulae give
\begin{subequations}
\label{EuclideanMixedReductions}
\begin{align}
\label{EuclideanMixedRankThree}
\overline{\chi^1}\gamma_{(3)}\chi^2
&=-S_3^{-1}K\wedge T,\\*
\label{EuclideanMixedRankFour}
\overline{\chi^1}\gamma_{(4)}\chi^2
&= - \mathrm{i} \,\star\bigl(\overline{\chi^1}\gamma_{(3)}\chi^2\bigr)
=-\frac{1}{2S_3}T\wedge T.
\end{align}
\end{subequations}
The factor $-\mathrm{i}$ in the last equation
follows from eq.~\eqref{EuclideanCliffordDual}.
The corresponding Lorentzian Hodge relation is the
gauged active class case of eq.~(A.14.11) of \cite{Levi:2026fpa}.
Thus no independent 3-form bilinear is required in the gauged active class 
also in the Euclidean theory.

Using eq.~\eqref{EuclideanCliffordDual}, the Euclidean versions
of the Hodge relations in eq.~(A.14.8) of \cite{Levi:2026fpa} read
\begin{subequations}
\label{EuclideanActiveHodge}
\begin{align}
\label{EuclideanHodgeTwoForm}
\star T&=-\frac{\mathrm{i}}{2S_3^2}K\wedge T\wedge T,\\*
\label{EuclideanHodgeVector}
\star K&=\frac{\mathrm{i}}{6S_3^2}T\wedge T\wedge T.
\end{align}
\end{subequations}
In particular, the imaginary coefficients in these equations arise from 
eqs.~\eqref{EuclideanCliffordVolume} and \eqref{EuclideanOrientation}.

\paragraph{Algebraic and differential relations.}
Using eqs.~\eqref{EuclideanMixedReductions}, 
we get the scalar and 1-form algebraic relations 
\begin{subequations}
\label{EuclideanAlgebraicProjections}
\begin{align}
\label{EuclideanAlgebraicScalarProjection}
0&=4gS_3(X^2-X^{-3})+T\lrcorner F
-\frac{X^3}{2S_3}(T\wedge T)\lrcorner G,\\*
\label{EuclideanAlgebraicVectorProjection}
0&=5S_3\dd X+\iota_KF
-\frac{X^3}{S_3}T\lrcorner(\iota_KG),
\end{align}
\end{subequations}
where the contraction convention is that of eq.~(A.3.12)
of \cite{Levi:2026fpa}.
This is the Euclidean form of eqs.~(4.4a), (4.5b) of \cite{Levi:2026fpa}
needed here.

Let $\mathcal A^\pm$ denote the KSE operators in the equal-charge
specialization of eq.~\eqref{EuclideanAlgebraicSym},
acting on the spinor components with Cartan eigenvalues $(\pm,\pm)$.
Thus 
\be 
\mathcal A^+\chi^1=\mathcal A^-\chi^2=0.
\ee
Subtracting $\overline{\chi^2}\mathcal A^+\chi^1=0$ from
$\overline{\chi^1}\mathcal A^-\chi^2=0$ isolates the dilaton gradient, 
and gives eq.~\eqref{EuclidTransverseDilaton}, 
as in eq.~(4.4b) of \cite{Levi:2026fpa}.
Differentiating eq.~\eqref{EuclideanActiveScalars} and
substituting the equal-charge specialization of eq.~\eqref{EuclideanGravitino} 
and its Majorana conjugate from Appendix~\ref{EuclideanConjugateKSEs} gives
eqs.~\eqref{EuclideanChargedScalarEquations}
and \eqref{EuclideanDifferentialScalar},
the counterparts of eqs.~(4.50), (4.61) of \cite{Levi:2026fpa}.

The same substitution in $\nabla_\mu K_\nu$, together
with eq.~\eqref{EuclidTransverseDilaton}, gives
\begin{align}
\label{EuclideanVectorDerivative}
X^2\nabla_\mu K_\nu
={}&-2X\partial_{[\mu}XK_{\nu]}-XS_3F_{\mu\nu}
-\frac{\mathrm{i}}{2}X^4(\iota_K\star G)_{\mu\nu}
+gX^{-2}T_{\mu\nu}.
\end{align}
This is the Euclidean counterpart of eq.~(4.69) of
\cite{Levi:2026fpa}. Its symmetric and antisymmetric parts give
eqs.~\eqref{EuclidKillingVector} and
\eqref{EuclideanDifferentialVector}, respectively.

Finally, differentiating the 2-form bilinear gives the following relation using 
the algebraic relations:
\begin{align}
\label{EuclideanTwoFormKSECancellation}
\big[\dd(X^{-2}T)+\iota_KG\big]_{\mu\nu\rho}
=-\frac{2\mathrm{i}}{X^2}\big[
\overline{\mathcal A^+\chi^1}\gamma_{\mu\nu\rho}\chi^2
-\overline{\chi^1}\gamma_{\mu\nu\rho}\mathcal A^-\chi^2
\big]=0, 
\end{align}
which establishes eq.~\eqref{EuclideanDifferentialTwoForm}.

\subsection{Higher-Rank Differential Bilinear Relations}
\label{EuclideanActiveHigherIdentities}

\paragraph{Rank 4.}
To derive the required rank-4 differential relation, we first use
eq.~\eqref{EuclideanTwoFormEndomorphism} to obtain, for a transverse
3-form $\beta$,
\be
\label{EuclideanTransverseThreeFormIdentity}
(T\lrcorner\beta)\wedge T\wedge T
=2S_3^2T\wedge\beta,
\qquad \iota_K\beta=0,
\ee
which is associated with the contraction identity in
eq.~(A.13.4a) of \cite{Levi:2026fpa}.
Using eqs.~\eqref{EuclideanDifferentialScalar} and
\eqref{EuclideanDifferentialTwoForm}, we then find
\begin{align}
\label{EuclideanWeightedFourFormDerivative}
\dd(S_3^{-1}T\wedge T)
={}&\frac{1}{XS_3^2}
\big[(5S_3\dd X+\iota_KF)\wedge T\wedge T
-2X^3S_3T\wedge\iota_KG\big]=0,
\end{align}
where the last equality uses
eqs.~\eqref{EuclideanAlgebraicVectorProjection}, 
\eqref{EuclideanTransverseThreeFormIdentity}.
In the notation $S_3=-ch$, $c^2=1$, $T=hJ$  of
eqs.~(3.39b), (3.43), (4.58) of \cite{Levi:2026fpa},
eq.~\eqref{EuclideanWeightedFourFormDerivative} is
$\dd(hJ\wedge J)=0$ of the gauged active class, 
as stated in Section~5.1.4 there.

\paragraph{Rank 5.}
We also need a rank-5 relation, for which we take only the transverse part
$F_\perp=F+S_3^{-2}K\wedge\iota_KF$. 
The geometric and Hodge relations in eqs.~\eqref{EuclideanTwoFormEndomorphism} 
and \eqref{EuclideanActiveHodge} imply
\begin{subequations}
\label{EuclideanTraceWedgeIdentities}
\begin{align}
\label{EuclideanTwoFormTraceWedge}
F_\perp\wedge T\wedge T
&=\frac{T\lrcorner F}{3S_3^2}T\wedge T\wedge T,\\*
\label{EuclideanFourFormTraceWedge}
(\iota_K\star G)\wedge T\wedge T
&=\big[(T\wedge T)\lrcorner G\big]\star K.
\end{align}
\end{subequations}
Differentiating eq.~\eqref{EuclideanHodgeTwoForm}, and using
eqs.~\eqref{EuclideanDifferentialScalar}, \eqref{EuclideanDifferentialVector},
\eqref{EuclideanWeightedFourFormDerivative}, 
\eqref{EuclideanTraceWedgeIdentities}, gives
\begin{align}
\label{EuclideanFiveFormIdntity}
\dd(X\star T)
={}&\frac{\mathrm{i}}{2S_3}F\wedge T\wedge T
+\left[
\frac{T\lrcorner F}{S_3}
-\frac{X^3}{2S_3^2}(T\wedge T)\lrcorner G
-6gX^{-3}\right]\star K \nn\\
={}&\frac{\mathrm{i}}{2S_3}F\wedge T\wedge T
 - 2 g \left(X^{-3}+2X^2\right)\star K,
\end{align}
where eq.~\eqref{EuclideanAlgebraicScalarProjection} 
was used for the second equality.

\paragraph{Rank 3.}
For the global equivariant polyform of the gauge-invariant action 
in Section~\ref{RelativeClass}, we also
need the differential relation of the 3-form already reduced
in eq.~\eqref{EuclideanMixedRankThree}. 
We take $G_\perp=G+S_3^{-2}K\wedge\iota_KG$, analogously to $F_\perp$ above.
The scalar projection in
eq.~\eqref{EuclideanAlgebraicScalarProjection} and the transverse
rank-2 projection of the algebraic KSE give the 4-form relation
\begin{align}
\label{EuclideanNeutralAlgebraicIdentity}
X^{-2}\iota_K\star F
={}&-\mathrm{i}X^{-2}F_\perp\wedge T
+\frac{X}{S_3}(\iota_K\star G)\wedge T
+ \mathrm{i} XS_3G_\perp
+\frac{2\mathrm{i}g}{S_3}(X^{-5}-1)T\wedge T.
\end{align}
Differentiating $K\wedge T/(XS_3)$, using
eqs.~\eqref{EuclideanDifferentialBilinears}, and applying
eq.~\eqref{EuclideanNeutralAlgebraicIdentity}, we then get
\begin{align}
\label{EuclideanNeutralThreeFormDerivative}
\dd\left(\frac{1}{XS_3}K\wedge T\right)
={}&-\mathrm{i}X^{-2}\iota_K\star F
-X^{-2}F\wedge T-XS_3G
+\frac{2g}{S_3}T\wedge T.
\end{align}
The transverse part follows from the algebraic identity 
in eq.~\eqref{EuclideanNeutralAlgebraicIdentity}, and contracting with $K$ 
verifies the longitudinal part directly from 
eqs.~\eqref{EuclideanDifferentialScalar}, \eqref{EuclideanDifferentialTwoForm}.

\bibliographystyle{JHEPmod}
\bibliography{equivar}

\providecommand{\href}[2]{#2}\begingroup\raggedright\begin{thebibliography}{10}

\bibitem{Maldacena:1997re}
J.~M. Maldacena, \emph{{The Large $N$ limit of superconformal field theories
  and supergravity}},
  \href{https://doi.org/10.4310/ATMP.1998.v2.n2.a1}{\emph{Adv. Theor. Math.
  Phys.} {\bfseries 2} (1998) 231}
  [\href{https://arxiv.org/abs/hep-th/9711200}{{\ttfamily hep-th/9711200}}].

\bibitem{Nahm:1977tg}
W.~Nahm, \emph{{Supersymmetries and Their Representations}},
  \href{https://doi.org/10.1016/0550-3213(78)90218-3}{\emph{Nucl. Phys. B}
  {\bfseries 135} (1978) 149}.

\bibitem{Nastase:1999cb}
H.~Nastase, D.~Vaman and P.~van Nieuwenhuizen, \emph{{Consistent nonlinear
  Kaluza--Klein reduction of 11d supergravity on $AdS_7\times S^4$ and
  self-duality in odd dimensions}},
  \href{https://doi.org/10.1016/S0370-2693(99)01266-6}{\emph{Phys. Lett. B}
  {\bfseries 469} (1999) 96}
  [\href{https://arxiv.org/abs/hep-th/9905075}{{\ttfamily hep-th/9905075}}].

\bibitem{Dibitetto:2017tve}
G.~Dibitetto and N.~Petri, \emph{{BPS objects in D = 7 supergravity and their
  M-theory origin}}, \href{https://doi.org/10.1007/JHEP12(2017)041}{\emph{JHEP}
  {\bfseries 12} (2017) 041}
  [\href{https://arxiv.org/abs/1707.06152}{{\ttfamily 1707.06152}}].

\bibitem{Conti:2024rwd}
A.~Conti, G.~Dibitetto, Y.~Lozano, N.~Petri and A.~Ram{\'\i}rez,
  \emph{{Half-BPS Janus solutions in AdS$_{7}$}},
  \href{https://doi.org/10.1007/JHEP12(2024)198}{\emph{JHEP} {\bfseries 12}
  (2024) 198} [\href{https://arxiv.org/abs/2407.21619}{{\ttfamily
  2407.21619}}].

\bibitem{Bobev:2023bxl}
N.~Bobev, M.~David, J.~Hong and R.~Mouland, \emph{{AdS$_{7}$ black holes from
  rotating M5-branes}},
  \href{https://doi.org/10.1007/JHEP09(2023)143}{\emph{JHEP} {\bfseries 09}
  (2023) 143} [\href{https://arxiv.org/abs/2307.06364}{{\ttfamily
  2307.06364}}]. {\textit{Erratum: ibid.} \textbf{09} (2023) 198
  [\href{https://doi.org/10.1007/JHEP09(2023)198}{DOI}]}.

\bibitem{Bobev:2025xan}
N.~Bobev, M.~David, V.~Dimitrov and J.~Hong, \emph{{AdS$_{7}$ black holes and
  holography}}, \href{https://doi.org/10.1007/JHEP11(2025)159}{\emph{JHEP}
  {\bfseries 11} (2025) 159}
  [\href{https://arxiv.org/abs/2509.02833}{{\ttfamily 2509.02833}}].

\bibitem{Hristov:2019mqp}
K.~Hristov, S.~Katmadas and C.~Toldo, \emph{{Matter-coupled supersymmetric
  Kerr-Newman-AdS$_4$ black holes}},
  \href{https://doi.org/10.1103/PhysRevD.100.066016}{\emph{Phys. Rev. D}
  {\bfseries 100} (2019) 066016}
  [\href{https://arxiv.org/abs/1907.05192}{{\ttfamily 1907.05192}}].

\bibitem{Kunduri:2006ek}
H.~K. Kunduri, J.~Lucietti and H.~S. Reall, \emph{{Supersymmetric multi-charge
  AdS$_5$ black holes}},
  \href{https://doi.org/10.1088/1126-6708/2006/04/036}{\emph{JHEP} {\bfseries
  04} (2006) 036} [\href{https://arxiv.org/abs/hep-th/0601156}{{\ttfamily
  hep-th/0601156}}].

\bibitem{Levi:2026fpa}
M.~Levi, \emph{{Timelike Supersymmetric Geometries in Seven-Dimensional
  Supergravity}},  \href{https://arxiv.org/abs/2610.31191}{{\ttfamily
  2610.31191}}.

\bibitem{Freedman:2012zz}
D.~Z. Freedman and A.~Van~Proeyen, \emph{{Supergravity}}. Cambridge Univ.
  Press, Cambridge, UK, 2012,
  \href{https://doi.org/10.1017/CBO9781139026833}{10.1017/CBO9781139026833}.

\bibitem{BenettiGenolini:2023kxp}
P.~Benetti~Genolini, J.~P. Gauntlett and J.~Sparks, \emph{{Equivariant
  Localization in Supergravity}},
  \href{https://doi.org/10.1103/PhysRevLett.131.121602}{\emph{Phys. Rev. Lett.}
  {\bfseries 131} (2023) 121602}
  [\href{https://arxiv.org/abs/2306.03868}{{\ttfamily 2306.03868}}].

\bibitem{Martelli:2023oqk}
D.~Martelli and A.~Zaffaroni, \emph{{Equivariant localization and holography}},
  \href{https://doi.org/10.1007/s11005-023-01752-1}{\emph{Lett. Math. Phys.}
  {\bfseries 114} (2024) 15}
  [\href{https://arxiv.org/abs/2306.03891}{{\ttfamily 2306.03891}}].

\bibitem{BenettiGenolini:2023ndb}
P.~Benetti~Genolini, J.~P. Gauntlett and J.~Sparks, \emph{{Equivariant
  localization for AdS/CFT}},
  \href{https://doi.org/10.1007/JHEP02(2024)015}{\emph{JHEP} {\bfseries 02}
  (2024) 015} [\href{https://arxiv.org/abs/2308.11701}{{\ttfamily
  2308.11701}}].

\bibitem{Colombo:2023fhu}
E.~Colombo, F.~Faedo, D.~Martelli and A.~Zaffaroni, \emph{{Equivariant volume
  extremization and holography}},
  \href{https://doi.org/10.1007/JHEP01(2024)095}{\emph{JHEP} {\bfseries 01}
  (2024) 095} [\href{https://arxiv.org/abs/2309.04425}{{\ttfamily
  2309.04425}}].

\bibitem{BerlineVergne}
N.~Berline and M.~Vergne, \emph{{Classes caract\'eristiques \'equivariantes.
  Formules de localisation en cohomologie \'equivariante}}, {\emph{C.R. Acad.
  Sc. Paris} {\bfseries 295} (1982) 539}.

\bibitem{Atiyah:1984px}
M.~F. Atiyah and R.~Bott, \emph{{The moment map and equivariant cohomology}},
  \href{https://doi.org/10.1016/0040-9383(84)90021-1}{\emph{Topology}
  {\bfseries 23} (1984) 1}.

\bibitem{Cassani:2024kjn}
D.~Cassani, A.~Ruip\'erez and E.~Turetta, \emph{{Localization of the 5D
  supergravity action and Euclidean saddles for the black hole index}},
  \href{https://doi.org/10.1007/JHEP12(2024)086}{\emph{JHEP} {\bfseries 12}
  (2024) 086} [\href{https://arxiv.org/abs/2409.01332}{{\ttfamily
  2409.01332}}].

\bibitem{Colombo:2025ihp}
E.~Colombo, V.~Dimitrov, D.~Martelli and A.~Zaffaroni, \emph{{Equivariant
  localization in supergravity in odd dimensions}},
  \href{https://arxiv.org/abs/2502.15624}{{\ttfamily 2502.15624}}.

\bibitem{Cassani:2025iix}
D.~Cassani, A.~Ruip{\'e}rez and E.~Turetta, \emph{{Bubbling saddles of the
  gravitational index}},
  \href{https://doi.org/10.21468/SciPostPhys.19.5.134}{\emph{SciPost Phys.}
  {\bfseries 19} (2025) 134}
  [\href{https://arxiv.org/abs/2507.12650}{{\ttfamily 2507.12650}}].

\bibitem{BenettiGenolini:2025icr}
P.~Benetti~Genolini, J.~P. Gauntlett, Y.~Jiao, J.~Park and J.~Sparks,
  \emph{{Equivariant localization for D = 5 gauged supergravity}},
  \href{https://doi.org/10.1007/JHEP03(2026)080}{\emph{JHEP} {\bfseries 03}
  (2026) 080} [\href{https://arxiv.org/abs/2508.08207}{{\ttfamily
  2508.08207}}].

\bibitem{Colombo:2025yqy}
E.~Colombo, V.~Dimitrov, D.~Martelli and A.~Zaffaroni, \emph{{Patch-wise
  localization with Chern-Simons forms in five dimensional supergravity}},
  \href{https://arxiv.org/abs/2511.13824}{{\ttfamily 2511.13824}}.

\bibitem{Cassani:2026teb}
D.~Cassani and E.~Turetta, \emph{{The black hole at the end of the cone:
  localizing the anomaly polynomial on toric geometries}},
  \href{https://arxiv.org/abs/2606.16986}{{\ttfamily 2606.16986}}.

\bibitem{Suh:2024asy}
M.~Suh, \emph{{Equivariant localization for wrapped M5-branes and D4-branes}},
  \href{https://doi.org/10.1007/JHEP09(2026)098}{\emph{JHEP} {\bfseries 09}
  (2026) 098} [\href{https://arxiv.org/abs/2404.01386}{{\ttfamily
  2404.01386}}].

\bibitem{BenettiGenolini:2026cdw}
P.~Benetti~Genolini, F.~Gaar, J.~P. Gauntlett, J.~Park and J.~Sparks,
  \emph{{Odd-Dimensional Localization in Supergravity}},
  \href{https://arxiv.org/abs/2607.07621}{{\ttfamily 2607.07621}}.

\bibitem{Freedman:2013oja}
D.~Z. Freedman and S.~S. Pufu, \emph{{The holography of $F$-maximization}},
  \href{https://doi.org/10.1007/JHEP03(2014)135}{\emph{JHEP} {\bfseries 03}
  (2014) 135} [\href{https://arxiv.org/abs/1302.7310}{{\ttfamily 1302.7310}}].

\bibitem{Kosmann:1971}
Y.~Kosmann, \emph{{D{\'e}riv{\'e}es de Lie des spineurs}},
  \href{https://doi.org/10.1007/BF02428822}{\emph{Ann. Mat. Pura Appl.}
  {\bfseries 91} (1971) 317}.

\bibitem{Berline:1983aa}
N.~Berline and M.~Vergne, \emph{{Zeros d’un champ de vecteurs et classes
  caracteristiques equivariantes}},
  \href{https://doi.org/10.1215/S0012-7094-83-05024-X}{\emph{Duke Mathematical
  Journal} {\bfseries 50} (1983) 539 }.

\bibitem{Ortin:2002qb}
T.~Ortin, \emph{{A Note on Lie-Lorentz derivatives}},
  \href{https://doi.org/10.1088/0264-9381/19/15/101}{\emph{Class. Quant. Grav.}
  {\bfseries 19} (2002) L143}
  [\href{https://arxiv.org/abs/hep-th/0206159}{{\ttfamily hep-th/0206159}}].

\bibitem{Couzens:2025ghx}
C.~Couzens, C.~M. Gregory, D.~Muniz, T.~Sieper and J.~Sparks, \emph{{Localizing
  Romans supergravity}},
  \href{https://doi.org/10.1007/JHEP09(2025)103}{\emph{JHEP} {\bfseries 09}
  (2025) 103} [\href{https://arxiv.org/abs/2504.07168}{{\ttfamily
  2504.07168}}].

\bibitem{BenettiGenolini:2024lbj}
P.~Benetti~Genolini, J.~P. Gauntlett, Y.~Jiao, A.~L{\"u}scher and J.~Sparks,
  \emph{{Equivariant localization for D = 4 gauged supergravity}},
  \href{https://doi.org/10.1007/JHEP08(2025)211}{\emph{JHEP} {\bfseries 08}
  (2025) 211} [\href{https://arxiv.org/abs/2412.07828}{{\ttfamily
  2412.07828}}].

\bibitem{Ferrero:2021etw}
P.~Ferrero, J.~P. Gauntlett and J.~Sparks, \emph{{Supersymmetric spindles}},
  \href{https://doi.org/10.1007/JHEP01(2022)102}{\emph{JHEP} {\bfseries 01}
  (2022) 102} [\href{https://arxiv.org/abs/2112.01543}{{\ttfamily
  2112.01543}}].

\bibitem{Hosseini:2017mds}
S.~M. Hosseini, K.~Hristov and A.~Zaffaroni, \emph{{An extremization principle
  for the entropy of rotating BPS black holes in AdS$_{5}$}},
  \href{https://doi.org/10.1007/JHEP07(2017)106}{\emph{JHEP} {\bfseries 07}
  (2017) 106} [\href{https://arxiv.org/abs/1705.05383}{{\ttfamily
  1705.05383}}].

\bibitem{Hosseini:2018dob}
S.~M. Hosseini, K.~Hristov and A.~Zaffaroni, \emph{{A note on the entropy of
  rotating BPS AdS$_7\times S^4$ black holes}},
  \href{https://doi.org/10.1007/JHEP05(2018)121}{\emph{JHEP} {\bfseries 05}
  (2018) 121} [\href{https://arxiv.org/abs/1803.07568}{{\ttfamily
  1803.07568}}].

\bibitem{Cabo-Bizet:2018ehj}
A.~Cabo-Bizet, D.~Cassani, D.~Martelli and S.~Murthy, \emph{{Microscopic origin
  of the Bekenstein-Hawking entropy of supersymmetric AdS$_{5}$ black holes}},
  \href{https://doi.org/10.1007/JHEP10(2019)062}{\emph{JHEP} {\bfseries 10}
  (2019) 062} [\href{https://arxiv.org/abs/1810.11442}{{\ttfamily
  1810.11442}}].

\bibitem{Cassani:2019mms}
D.~Cassani and L.~Papini, \emph{{The BPS limit of rotating AdS black hole
  thermodynamics}}, \href{https://doi.org/10.1007/JHEP09(2019)079}{\emph{JHEP}
  {\bfseries 09} (2019) 079}
  [\href{https://arxiv.org/abs/1906.10148}{{\ttfamily 1906.10148}}].

\bibitem{Choi:2018hmj}
S.~Choi, J.~Kim, S.~Kim and J.~Nahmgoong, \emph{{Large AdS black holes from
  QFT}}, \href{https://doi.org/10.1007/s40042-026-01671-z}{\emph{J. Korean
  Phys. Soc.} {\bfseries 89} (2026) 697}
  [\href{https://arxiv.org/abs/1810.12067}{{\ttfamily 1810.12067}}].

\bibitem{Nahmgoong:2019hko}
J.~Nahmgoong, \emph{{6d superconformal Cardy formulas}},
  \href{https://doi.org/10.1007/JHEP02(2021)092}{\emph{JHEP} {\bfseries 02}
  (2021) 092} [\href{https://arxiv.org/abs/1907.12582}{{\ttfamily
  1907.12582}}].

\bibitem{Bobev:2015kza}
N.~Bobev, M.~Bullimore and H.-C. Kim, \emph{{Supersymmetric Casimir Energy and
  the Anomaly Polynomial}},
  \href{https://doi.org/10.1007/JHEP09(2015)142}{\emph{JHEP} {\bfseries 09}
  (2015) 142} [\href{https://arxiv.org/abs/1507.08553}{{\ttfamily
  1507.08553}}].

\bibitem{Ohmori:2021dzb}
K.~Ohmori and L.~Tizzano, \emph{{Anomaly matching across dimensions and
  supersymmetric Cardy formulae}},
  \href{https://doi.org/10.1007/JHEP12(2022)027}{\emph{JHEP} {\bfseries 12}
  (2022) 027} [\href{https://arxiv.org/abs/2112.13445}{{\ttfamily
  2112.13445}}].

\end{thebibliography}\endgroup

\end{document}